\documentclass[12pt]{article}
\usepackage{graphics}
\usepackage{graphicx}
\DeclareGraphicsExtensions{.pdf}
\usepackage{float} 
\usepackage{amsmath}
\usepackage{slashed}
\usepackage{amsfonts}   
\usepackage{amssymb}

\begin{document}
\date{}
\title{{\bf{\Large Holographic duals of $ \mathcal{N}=1 $ quivers in 5d and their nonrelativistic limits}}}
\author{
 {\bf {\normalsize Dibakar Roychowdhury}$
$\thanks{E-mail:  dibakarphys@gmail.com, dibakar.roychowdhury@ph.iitr.ac.in}}\\
 {\normalsize  Department of Physics, Indian Institute of Technology Roorkee,}\\
  {\normalsize Roorkee 247667, Uttarakhand, India}
\\[0.3cm]
}

\maketitle
\begin{abstract}
We explore nonrelativistic limits of $\mathcal{N}=1$ quiver gauge theories in 5d. The stringy counterpart of these SCFTs is characterised by torsional string Newton-Cartan (TSNC) sigma models those are defined over non-Lorentzian manifolds. We further show that under transverse T-duality, these TSNC sigma models are mapped into another new class of nonrelativistic sigma models those are defined over a T-dual TSNC background. Considering nonrelativistic limits of various field theory observables in a holographic set up, we further estimate corresponding entities in the TSNC limit of $ \mathcal{N}=1 $ quivers. We carry out a parallel analysis on holomorphic functions and the associated pole structures in the nonrelativistic limit of ($ p ,q $) five brane webs. In particular, we investigate the generic structure of various loop operators in a nonrelativistic set up and explore their properties under S-duality. Finally, we comment on the large $ c $ limit of RR fields and discuss the associated S-duality transformation rules in the nonrelativistic limit of $ \mathcal{N}=1 $ quivers.
\end{abstract}
\section{Introduction and summary}
\subsection{Introduction and the general idea of this paper}
Nonrelativistic string theory \cite{Gomis:2000bd}-\cite{Gomis:2005pg} plays a pivotal role in understanding the nonrelativistic limits of classical gravity \cite{Bergshoeff:2015uaa}-\cite{Bergshoeff:2019pij} as well as the limits of gauge/string duality \cite{Bergshoeff:2018yvt}-\cite{Harmark:2019upf}. Besides, it opens up a window for a profound understanding of the quantum gravity in its nonrelativistic limits \cite{Gallegos:2019icg}-\cite{Gomis:2019zyu}.

Recently, there has been a significant progress in understanding the nonrelativistic target space geometries for the NS-NS sector of the closed bosonic strings \cite{Bidussi:2021ujm}-\cite{Yan:2021lbe}. These backgrounds, which we term as the torsional String-Newton Cartan (TSNC) backgrounds, are characterised by the following set of data \cite{Bidussi:2021ujm}
\begin{eqnarray}
\tau_{\mu}^A ~~;~~ h_{\mu \nu}=e_{\mu}^a e_{\nu}^b\delta_{ab}~~;~~m_{\mu \nu}=\eta_{AB}\tau^A_{[\mu}\pi_{\nu]}^B +\delta_{ab}e_{[\mu}^a\pi_{\nu]}^b
\end{eqnarray}
where $ \tau_{\mu}^A ~(A=0,1)$ are cloak one forms associated with TSNC target space, $ h_{\mu \nu} $ is the metric of the transverse space and $ m_{\mu \nu} $ is the two form that couples with the tension current. 

The transverse gauge fields $ \pi^{a}_{\mu} ~(a=2,\cdots , d-1)$ are what we identify as the key elements of the TSNC (or the nonrelativistic geometric) data set. These characterise the transverse $ B_{\mu \nu} $ field of the nonrelativistic target space. At the level of the algebra, they introduce an additional global charge $ Q_a $ that result in a F-string Galilei algebra  \cite{Bidussi:2021ujm}.

One of the prime motivations of the present analysis is to understand these nonrelativistic geometric data from the perspective of the gauge/string correspondence.
There exists a plethora of examples of holoraphic dualities those go beyond the celebrated correspondence between type IIB superstrings in $ AdS_5 \times S^5 $ and $ \mathcal{N}=4 $ SYM in 4d \cite{Maldacena:1997re}. It is therefore natural to ask how to take a consistent nonrelativistic limit in any of these examples and how could one make sense of the holographic correspondence in these (TSNC) limits. 

The purpose of this paper is to shed light on some of these isssues and set stage for a deeper understanding on nonrelativistic holography going beyond the standard Maldacena conjecture \cite{Maldacena:1997re}. To address these issues, we take the specific example of $ \mathcal{N}=1 $ dualities those relate type IIB string theory in $ AdS_6 \times S^2 \times \Sigma_{(2)} $ and $ \mathcal{N}=1 $ SCFTs \cite{Seiberg:1996bd} in 5d. 

The present paper heavily relies on the two parallel holographic descriptions of $ \mathcal{N}=1 $ quivers in 5d- (i) the DGKU solution those were originally developed by authors in \cite{DHoker:2016ujz}-\cite{DHoker:2017zwj} and subsequently extended in \cite{Fluder:2020pym}-\cite{Bergman:2018hin} and (ii) the electrostatic description introduced by authors in \cite{Legramandi:2021uds} and subsequently explored in \cite{Roychowdhury:2021jqt} in the context of classical integrability.

The DGKU solution \cite{DHoker:2016ujz}-\cite{DHoker:2017zwj} was originally proposed as a warped product of $ AdS_6 \times S^2 $ over the 2d Riemann surface parametrised by a set of complex coordinates. The full type IIB solution is characterized in terms of a pair of locally holomorphic functions ($ \mathcal{A}_{\pm}(\omega) $) defined over the 2d Riemann surface. The poles associated with the differential of these holomorphic functions correspond to the locations of the ($ p ,q $) five branes along the real line when the Riemann surface is considered to be the upper half of the complex plane. The charges associated with the five brane web is given by the residues at these poles.

The electrostatic approach of \cite{Legramandi:2021uds}, on the other hand, considers an intersection of NS5-D5-D7 brane configuration in 10d. The NS5 branes are placed at discrete locations along the holographic ($ \eta $) axis while the color D5 branes are stretched in between them. The flavour nodes of the quiver are sourced due to D7 branes. 

The spacing between these NS5 branes is what measures the strength of the coupling between the hypermultiplets and the vector multiplets of the dual SCFTs. As our analysis reveals, in the strict nonrelativistic limit, these NS5 branes are pushed on top of each other which results in a strongly coupled description for the dual quiver in its nonrelativistic limit.

In spite of these developments, some key issues are yet to be addressed. These are precisely the questions in the sense of the TSNC limits those are alluded to the above. The present paper discusses the physical consequences in these limits and how one could make sense of it in the context of the gauge/string duality. To be more precise, below we pose some of the key questions those motivate the present analysis of the paper. 

$ \bullet $ It is utmost important to understand whether and how the two seemingly different approaches (\cite{DHoker:2016ujz}-\cite{DHoker:2017zwj} and \cite{Legramandi:2021uds}) to 5d $ \mathcal{N}=1 $ quivers can lead to identical physical phenomena in their respective nonrelativistic limits. For example, whether one can still identify the ($ p,q $) brane web as a pole associated with the differential of the holomorphic function in the large $ c $ limit. This must get translated into an equivalent picture of five brane web while considering a large $ c $ limit of Hanany-Witten like  brane set up as discussed in \cite{Legramandi:2021uds}.

$ \bullet $ What is the NS5-D5-D7 brane set up and the associated Page charges in the TSNC limit of the type IIB background? These numbers must agree to those obtained in the nonrelativistic limit and using a holomorphic function approach of \cite{DHoker:2016ujz}-\cite{DHoker:2017zwj}.

$ \bullet $ What does happen to the various physical observables (for example the central charge, couplings, Wilson loops etc.) associated with the dual QFT in its TSNC limit?

$ \bullet $ What is the fate of S-duality transformation rules in the nonrelativistic limits of $ \mathcal{N}=1 $ super-conformal quivers? 

We wish to gain some insights into these issues using the nonrelativistic stringy counterpart of the correspondence. These nonrelativistic sigma models, as we show, could be systematically obtained taking a TSNC limit of type IIB $ AdS_6 \times S^2 \times \Sigma_{(2)} $ background.

Before, we move on to the summary of results, it is customary to outline a few steps those realise these TSNC backgrounds as a solution of type IIB supergravity equations of motion in the nonrelativistic limit\footnote{For the purpose of the present paper, we focus only on the NS sector of the type IIB solution. The primary reason for this stems from the fact that our current understanding of the F- string Galilei generators and the associated gauge fields (which we identify as the TSNC data in this paper) is limited only to the NS sector \cite{Bidussi:2021ujm}. The inclusion of the RR sector would require further understanding on the F- string Galilei algebra for its full supersymmetric generalisation which is beyond the scope of the present analysis. In other words, the supersymmetric extension of the F- string Galilei algebra would require a geometric realisation for the RR sector which would give rise to an additonal set of RR generators and the associated gauge fields. These gauge fields would certainly add to the existing set of TSNC data thereby making it a supersymmetric background endowed with nonrelativistic symmetries.}. This goes precisely along the line of \cite{Bergshoeff:2021bmc}. Since, we have the TSNC data available only for the NS-NS sector of the full type IIB supergravity solutions, therefore following the discussion below, one could imagine taking a large $ c $ limit of the equations of motion in the NS sector only. 

TSNC limit corresponds to an expansion of the background fields of the form
\begin{eqnarray}
\label{Eq1.2}
\mathcal{G}_{\mu \nu}=c^2 E^A_{\mu}E^B_{\nu}\eta_{AB}+\delta_{ab}e^a_{\mu}e^b_{\nu}~;~B_{\mu \nu}=c^2 \eta_{AB}E^A_{[\mu}\Pi^B_{\nu]}+\delta_{ab}e^a_{[ \mu}\pi^b_{\nu]},
\end{eqnarray}
where we define \cite{Bidussi:2021ujm}
\begin{eqnarray}
\Pi^A_{\mu} = \epsilon^A_B \tau_{\mu}^B +\frac{1}{2c^2}\pi_{\mu}^A,
\end{eqnarray}
as the longitudinal component of the background $ \pi_{\mu} $ gauge fields.

On the other hand 
\begin{eqnarray}
\label{Eq1.4}
E^A_{\mu} = \tau_{\mu}^A +\frac{1}{2c^2}\pi_{\mu}^B \epsilon_B^A ,
\end{eqnarray}
are the longitudinal components of the background vielbeins of the relativistic spacetime.

Using (\ref{Eq1.2})-(\ref{Eq1.4}), one could expand the curvature two form ($ \mathcal{R} $) and the NS-NS three form ($ \mathcal{H}^{(3)} $) in the large $ c $ limit. Upon substituting back these into the type IIB action, one ends up in a large $ c $ expansion of the action \cite{Bergshoeff:2021bmc}. The equations of motion for TSNC data are readily obtained by varying the zeroth order (finite) action $\mathcal{S}^{(0)}$.
\subsection{Summary of results}
Below, we summarise the key findings of the present paper. At the first place, we list those results/observations considering the electrostatic description \cite{Legramandi:2021uds} of $ \mathcal{N}=1 $ quivers.

$ \bullet $ The TSNC scaling results in a picture of \emph{collapsing} branes at the origin of the holographic ($ \eta $) axis (see Fig.\ref{hw2}(a)). This has an effect in producing a metric singularity near the origin ($\sigma \sim 0 , \eta \sim 0 $) of the $ (\sigma , \eta) $ plane which also gets reflected in the corresponding TSNC data. In the holomorphic language of \cite{DHoker:2016ujz}-\cite{DHoker:2017zwj}, this is precisely translated into the picture of collapsing five brane web at the zero pole in the complex plane.

$ \bullet $ We calculate the number of NS5 branes as well as the color D5 branes in the TSNC limit of $ \mathcal{N}=1 $ quivers. These numbers, which are also called the Page charges, precisely match our expectations while calculating the residues (using the holomorphic functions of \cite{Bergman:2018hin}) at the zero pole of the nonrelativistic ($ p , q $) five brane web. It turns out that in both descriptions, the total number of branes is conserved while taking the large $ c \rightarrow \infty $ limit.

$ \bullet $ Transverse T-duality allows us to map these TSNC sigma models to another class of nonrelativistic sigma models \cite{Bergshoeff:2018yvt} those are propagating over T-dual TSNC manifold. We identify these T-duality rules and explicitly work them out taking specific example(s).

$ \bullet $ A further analysis on the QFT observables in the TSNC limit reveals a number of interesting facts. For example, the central charge in the nonrelativistic limit of $ \mathcal{N}=1 $ quivers goes with different powers of the number of the NS5 branes
\begin{eqnarray}
\hat{c}_{nr}\sim \begin{cases}
      N^2_c P^3&  (T_{N_c ,P}~\text{Quiver})\\
      N^2_c P^2 & (+_{P, N_c}~\text{Quiver})
    \end{cases} 
\end{eqnarray}
where $ P = \tilde{Q}_{NS5} $ is the number of NS5 branes in the nonrelativistic limit. 

$ \bullet $ The coupling between the tensor multiplet and the vector multiplets grows strong in the strict large $ c $ limit namely, $ g^2_{QFT} \sim c^2 \rightarrow \infty$. This indicates the onset of a strongly coupled dynamics in the nonrelativistic limit of $ \mathcal{N}=1 $ quivers. We identify this strong coupling behavior as an artefact of the collapsing NS5 branes in the nonrelativistic limit.

In section \ref{sec5}, we revisit the nonrelativistic limits of $ \mathcal{N}=1 $  quivers using the language of the locally holomorphic functions $ (\mathcal{A}_{\pm}(\omega)) $ \cite{DHoker:2016ujz}-\cite{DHoker:2017zwj}. We consider both the examples of balanced ($ T_{N_c} $ and $ +_{N_c , M} $) as well as unbalanced ($ Y_{N_c} $ and $ X_{N_c} $) quivers. 

$ \bullet $ In the case of balanced ($ T_{N_c} $ and $ +_{N_c ,M} $) quivers, we show that the differential of the holomorphic function ($ \partial_{\omega}\mathcal{A}_{\pm} $) exhibits a zero pole in the nonrelativistic limit of the supergravity solutions. The pole essentially corresponds to the origin of the complex coordinate system associated with the 2d Riemann surface ($ \Sigma_{(2)} $) and is lying along the real line when considering $ \Sigma_{(2)} $ to be the upper half plane. 

The existence of such zero pole corresponds to a description of collapsing ($ p ,q $) five brane web in the nonrelativistic limit of the $ \mathcal{N}=1 $ quiver.  We further show that the $ S^2 $ (of the internal space) shrinks to zero along the boundary of the 2d Riemann surface and ensures the geodesic completeness \cite{DHoker:2017mds} of the TSNC background.

$ \bullet $ The primary question, that we address, concerns about the fate of the S-duality transformation rules in the nonrelativistic ($ c\rightarrow \infty $) limit. In what follows, we compute various $ (p , q) $  loop operators in the nonrelativistic limit and and study their properties under S-duality transformations. We show that the S-duality transforms $ (p , q) $ string charges in such way so that $ (1,0) $ and $ (0,1) $ states are precisely exchanged in the dual description. 

$ \bullet $  We further compute antisymmetric Wilson loops using D3 brane embeddings and subsequently consider their nonrelativistic limits. We show that in the nonrelativistic limits of the ($ p ,q $) five brane web, the expectation values for these Wilson loops (at any point ($ \tilde{\omega} $) of the upper half plane) turn out to be $\langle \mathcal{W}(\tilde{z}) \rangle|_{c \rightarrow \infty} \simeq 1 - \mathcal{O}(c^{-5})  $. 

These results also agree with the results those were obtained using the electrostatic description of $ \mathcal{N}=1 $ quivers. We further estimate the associated F1 and D1 charges of the nonrelativistic D3 branes those are embedded at any point ($ \tilde{\omega} $) of the complex upper half plane and derive an identity analogous to those obtained previously in \cite{Uhlemann:2020bek}.

$ \bullet $ In sections \ref{sec 5.4} and \ref{sec5.5}, we present an algorithm in order to obtain nonrelativistic limits of unbalanced ($ Y_{N_c} $ and $ X_{N_c} $) quivers. Like in the case of balanced quivers, we identify a zero pole structure (at the origin of the complex coordinate system) associated with the derivatives of these holomorphic functions. We also discuss ($ p ,q $) loop operators in the nonrelativistic limits of unbalanced quivers and study their properties under S-duality.

We draw our conclusion in section \ref{sec6} and outline some possible future directions.

In Appendix \ref{appen A}, we consider large $ c $ limits of background RR-fluxes and the dilaton of the NS sector. We show that, under an S-duality, these solutions can be mapped into those of the nonrelativistic background solutions of \cite{Apruzzi:2018cvq}. We also discuss the nonrelativistic limits of matching solutions \cite{Legramandi:2021uds} those map the electrostatic variables into the DGKU solution \cite{DHoker:2016ujz}.
\section{Electrostatic description and TSNC limit}
\subsection{$ \mathcal{N}=1 $ backgrounds}
$ \mathcal{N}=1 $ quiver gauge theories in 5d have their dual description in terms of type IIB supergravity with an $ AdS_6 $ factor. The full 10d solution is expressed as warped product of the form $ AdS_6 \times S^2 \times \Sigma_{(2)} $. Here, $  \Sigma_{(2)} $ is a two dimensional Riemann surface parametrised by complex coordinates \cite{DHoker:2016ujz}-\cite{DHoker:2017zwj}.

In the original construction of the duality \cite{DHoker:2016ujz}-\cite{DHoker:2017zwj}, the warped factors of both $ AdS_6 $ and $ S^2 $ are expressed in terms of locally holomorphic functions of $ \Sigma_{(2)} $ which we dentify here as the DGKU solution. As we progress, along the way, we outline the roadmap to obtain the TSNC data corresponding to these solutions using the map of \cite{Legramandi:2021uds}.

The first part of the analysis is heavily based on the electrostatic viewpoint of $ \mathcal{N}=1 $ SCFTs as elaborated by authors in \cite{Legramandi:2021uds}. The 10d background contains an $ AdS_6 $ factor together with in internal manifold ($ \mathcal{M}_4 $) that comprises of an $ S^2 $ preserving the $ SU(2)_R $ symmetry of the dual SCFTs.

In the electrostatic description, one expresses the type IIB background in terms of a potential function $ V(\sigma , \eta) $ that satisfies Laplace's equation 
\begin{eqnarray}
\partial_{\sigma}(\sigma^2 \partial_{\sigma}V)+\sigma^2 \partial^2_{\eta}V=0,
\end{eqnarray}
with appropriate boundary conditions \cite{Legramandi:2021uds}
\begin{eqnarray}
\hat{V} (\sigma \rightarrow \pm \infty , \eta ) =0~~;~~\mathcal{R} (\eta =0) =0=\mathcal{R} (\eta =P),
\end{eqnarray}
where $ \mathcal{R} (\eta) $ is called the rank function that classifies different classes of quivers\footnote{Given the electrostatic description, one can imagine a charge distribution $ \mathcal{R} (\eta)  $ \cite{Legramandi:2021uds} between the conducting plates located at $ \eta =0 $ and $ \eta=P $. }. Here, the modified potential function $ \hat{V} = \sigma V $ (that actually solves the Laplace's equation) is subjected to the boundary conditions of the following form
\begin{eqnarray}
\hat{V} (\sigma , \eta =0 ) =0=\hat{V} (\sigma , \eta =P ),
\end{eqnarray}
where $ \eta $ is called the \emph{holographic} direction whose range is bounded between $ 0 $ and $ P $.

Using the string frame, the type IIB background could be formally expressed as \cite{Legramandi:2021uds}
\begin{eqnarray}
\label{e2.4}
ds^2_{IIB}&=&f_1 (\sigma , \eta) ds^2_{AdS_6}+ds^2_{\mathcal{M}_4}\\
&=& f_1 (\sigma , \eta) ds^2_{AdS_6}+f_2(\eta , \sigma)d\Omega_2 (\chi , \xi) +f_3 (\eta , \sigma)(d\sigma^2 +d\eta^2),\\
B_2 &=&f_4 (\sigma , \eta)\sin\chi d\chi \wedge d\xi ~~;~~\mathcal{C}_2 = f_5 (\sigma , \eta)\sin\chi d\chi \wedge d\xi ,\\
e^{-2\phi}&=&f_6(\sigma , \eta)~~;~~\mathcal{C}_0 = f_7 (\sigma , \eta),\\
f_1 &=&\frac{3\pi}{2}\left(\sigma^2 +\frac{3\dot{V}}{\partial^2_{\eta}V} \right)^{1/2}~;~f_2 = f_1\frac{\partial^2_{\eta}V \dot{V}}{3 \sigma \Delta}~;~ f_3 = f_1\frac{\partial^2_{\eta}V}{3\dot{V}},\\
f_4 &=&\frac{\pi}{2}\left(\eta -\frac{\dot{V}\partial_{\sigma}\partial_{\eta}V}{\Delta} \right) ~;~f_5 = \frac{\pi}{2}\left(V-\frac{\dot{V}}{\Delta}(\partial_{\eta}V(\partial_{\sigma}\partial_{\eta}V)-3\partial^2_{\eta}V \partial_{\sigma}V) \right),\\
f_6 &=& \frac{12\sigma \dot{V}\partial^2_{\eta}V \Delta}{(3\partial_{\sigma}V+\sigma \partial^2_{\eta}V)^2}~;~f_7 = 2 \left(\partial_{\eta}V+\frac{3\dot{V}\partial_{\sigma}\partial_{\eta}V}{(3\partial_{\sigma}V+\sigma \partial^2_{\eta}V)} \right), \\
\Delta &=& \frac{1}{\sigma}(2\dot{V}-\ddot{V})\partial^2_{\eta}V +\sigma (\partial_{\sigma}\partial_{\eta}V)^2~;~\dot{V}(\sigma , \eta)=\sigma \partial_{\sigma}V.
\label{e2.11}
\end{eqnarray}  

The corresponding Hanany-Witten set up consists of an intersection of NS5-D5-D7 branes in 10d. The NS5 branes are placed along the holographic $ \eta $ axis at discrete locations ($ \eta_i $) while the color D5 branes are extended between them. The completion of the $ \mathcal{N}=1 $ quiver is achieved by placing flavour D7 branes at $ \eta =P-1 $ (see Fig.\ref{hw1}).
\subsection{$  T_{N_c,P}$ quivers}
\begin{figure}
\includegraphics[scale=.85]{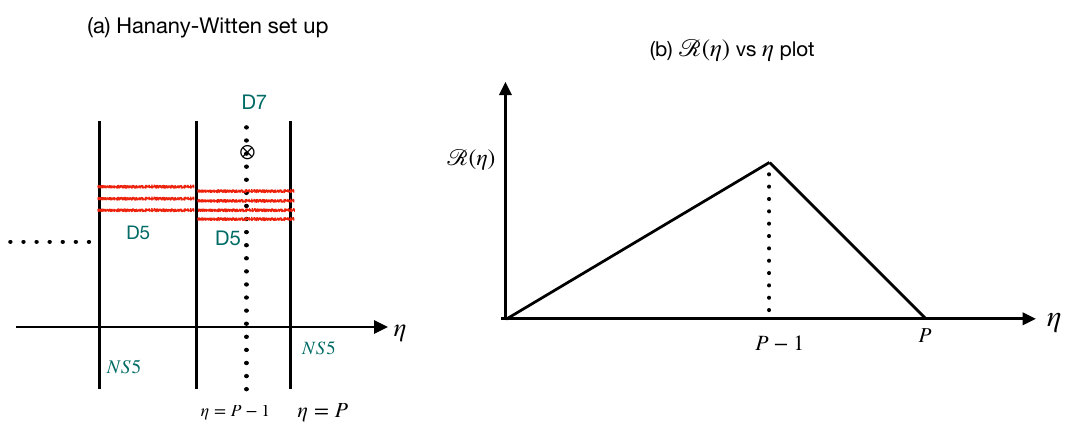}
  \caption{(a)NS5-D5-D7 brane intersections for $  T_{N_c,P}$ quivers in the electrostatic description of $ \mathcal{N}=1 $ SCFTs in 5d. (b) The corresponding rank function/chargge density $ \mathcal{R}(\eta) $ is plotted against the holographic $ \eta $ axis. The rank function increases linearly exhibiting a ``kink'' at the location of the flavour D7 branes.} \label{hw1}
\end{figure}

To start with, we first construct the nonrelativistic limit of $  T_{N_c,P}$ quivers which is closed by placing flavour branes at $ \eta =P-1 $ \cite{Legramandi:2021uds} (see Fig.\ref{hw1}). 

The corresponding rank function is given by
\begin{eqnarray}
\label{e2.12}
\mathcal{R}(\eta)= \begin{cases}
      N_c \eta &  0 \leq \eta \leq (P-1)\\
      N_c (P-1)(P- \eta) & (P-1)\leq \eta \leq P.
    \end{cases} 
\end{eqnarray}

Expanding the potential $ V(\sigma , \eta) $ in the limit of small $ \sigma $ and large $ P $ we find \cite{Roychowdhury:2021jqt}
\begin{eqnarray}
\label{e2.13}
\hat{V}(\sigma \sim 0 , \eta)\sim\frac{\eta N_c  P \log 2}{\pi }-\frac{\pi  \eta  \left(\eta ^2+1\right) N_c}{24 P}-\frac{\sigma \eta  N_c}{2} +\frac{\pi  \eta  \sigma ^2 N_c}{8 P}+\mathcal{O}(\sigma^3).
\end{eqnarray}

On the other hand, an expansion in the limit of large $ \sigma $ reveals 
\begin{eqnarray}
\label{2.14}
\hat{V}(\sigma \rightarrow \infty , \eta)\sim \frac{P^3 N_c }{\pi ^3}e^{-\frac{\pi  \sigma }{P}} \sin \left(\frac{\pi }{P}\right) \sin \left(\frac{\pi  \eta }{P}\right).
\end{eqnarray}

However, to carry out our analysis, we restrict ourselves in the region $ \sigma \sim 0 $ and consider\footnote{The rest of the directions of $ AdS_6 $ are being freezed.} $ (t , \eta) $ being the longitudinal directions of the TSNC manifold \cite{Bidussi:2021ujm}. The remaining directions of the internal manifold ($ \mathcal{M}_4 $) are being considered as the transverse coordinates.

The corresponding metric components read as 
\begin{eqnarray}
f_1 (\sigma \sim 0, \eta)&\sim &\frac{3\pi}{2}   \sqrt{\Big |-\frac{\eta ^2}{2}+\frac{12 P^2 \log 2}{\pi ^2}-\frac{1}{2}\Big |}+\mathcal{O}(\sigma^2) ~,\\
f_2 (\sigma \sim 0, \eta)&\sim &\frac{-3 \left(\pi ^2 \eta ^2 \left(24 P^2 \log 2-\pi ^2 \left(\eta ^2+1\right)\right)^{3/2}\right)}{\sqrt{2} \left(\pi ^4 \left(9 \eta ^4+12 \eta ^2-1\right)-576 P^4 \log ^22+48 \pi ^2 \left(1-6 \eta ^2\right) P^2 \log 2\right)},\\
f_3 (\sigma \sim 0, \eta)&\sim &\frac{3 \pi ^2}{\sqrt{|48 P^2 \log 2-2 \pi ^2 \left(\eta ^2+1\right)|}}+\mathcal{O}(\sigma^2).
\end{eqnarray}
\subsubsection{Decoding the vielbeins}
We are now in a position to decode the NR data $ \lbrace \tau_{\hat{\mu}}^A , h_{ij}, \pi_{i}^a \rbrace$ which we collectively call as the TSNC data \cite{Bidussi:2021ujm} associated with the non-Lorentzian manifold\footnote{As our analysis reveals, the longitudinal $ \pi_{\mu}^{A}~(A=0,1) $ gauge fields are identically zero for warped $ AdS_6 \times S^2 $ background considered in this paper. Following the redefinition of \cite{Bidussi:2021ujm}, the above feature clearly reflects the absence of the $ Z_A $ symmetry \cite{Bergshoeff:2021bmc} at the level of the TSNC sigma models those are discussed in this paper. This is an artefact of the non existence of the foliation/zero torsion constraint \cite{Bergshoeff:2021bmc},\cite{Bidussi:2021ujm} for TSNC backgrounds those are obtained as a limit of the type IIB supergravity solutions.}. Given these set of data, one can next write down the corresponding sigma model Lagrangian that is nonrelativistic with reference to the target space.

To obtain the TSNC data, we propose NR scaling of the following form
\begin{eqnarray}
\label{e2.18}
t = \tilde{t}~;~\eta = c^2 \tilde{\eta}~;~\xi = \tilde{\xi}~;~\chi =\frac{\tilde{\chi}}{c}.
\end{eqnarray}

Using (\ref{e2.18}), we find longitudinal vielbeins as\footnote{We remove tildes for simplicity which has been followed for the rest of the analysis.}
\begin{eqnarray}
\label{e2.19}
e_{t}^{\hat{0}}dt = c E^{0}_t dt = c\sqrt{\frac{3\pi \eta}{2\sqrt{2}}}\left( 1-\frac{6P^2 \log 2}{\pi^2 c^4 \eta^{2}}+ \cdots \right) dt.
\end{eqnarray}

Comparing (\ref{e2.19}) with \cite{Bidussi:2021ujm}
\begin{eqnarray}
E_t^0 = \tau_t^0 - \frac{1}{2c^2}\pi_t^1 ,
\end{eqnarray}
we find the corresponding TSNC data as
\begin{eqnarray}
\label{e2.21}
\tau_t^0 =  \sqrt{\frac{3\pi \eta}{2\sqrt{2}}} ~;~\pi_t^1 =0.
\end{eqnarray}

On a similar note, we find
\begin{eqnarray}
e_{\eta}^{\hat{1}}d\eta = c E^{1}_\eta d\eta =\frac{c\sqrt{3 \pi}}{(2\pi^2)^{1/4}\sqrt{\eta}}\left(1+\frac{6P^2 \log 2}{\pi^2 c^4 \eta^2} + \cdots \right),
\end{eqnarray}
which therefore yields
\begin{eqnarray}
\label{2.23}
\tau_{\eta}^1 = \frac{1}{(2\pi^2)^{1/4}}\sqrt{\frac{3 \pi}{\eta}}~;~\pi_{\eta}^0 =0.
\end{eqnarray}

To obtain the transverse vielbeins ($ e^a_{\mu}~,a=2,3 $), we notice
\begin{eqnarray}
e^{\hat{2}}_{\chi}d \chi &=&\sqrt{\frac{\pi \eta}{3\sqrt{2}}}d\chi = e^{2}_{\chi}d \chi ~;~\Rightarrow e^{2}_{\chi} = \sqrt{\frac{\pi \eta}{3\sqrt{2}}},\\
e^{\hat{3}}_{\xi}d \xi &=&\sqrt{\frac{\pi \eta}{3\sqrt{2}}}\chi d\xi =e^{3}_{\xi}d \xi ~;~\Rightarrow e^{3}_{\xi} =\sqrt{\frac{\pi \eta}{3\sqrt{2}}}\chi ,
\end{eqnarray}
where we take into account the $ \text{Im}\sqrt{f_2} $ while decoding the vielbeins.

The remaining TSNC data is obtained by looking into the transverse $ B_2 $ field and its NR reduction. A straightforward computation of the metric component $ f_4 (\sigma \sim 0, \eta)|_{P \gg 1} \sim \frac{\pi^3 \eta^3}{2 P^2 \log 2} $ \cite{Roychowdhury:2021jqt} reveals the following combination of the vielbein and the $ \pi $ fields \cite{Bidussi:2021ujm}
\begin{eqnarray}
\label{e2.26}
B^{(NR)}_{\chi \xi}=\frac{1}{2}(e_{\chi}^2\pi_{\xi}^2 -e_{\xi}^3 \pi_{\chi}^3) = \frac{\pi^3 \eta^3 \chi}{2 P^2 \log 2},
\end{eqnarray}
which is subjected to the scaling $ P \rightarrow c^2 P $ in accordance to that with the scaling of the holographic $ \eta $ axis (\ref{e2.18}).
\subsubsection{Sigma model and its symmetries}
Given the above set of TSNC data (\ref{e2.21})-(\ref{e2.26}), we are now in a position to write down the corresponding NR sigma model action \cite{Bidussi:2021ujm}
\begin{eqnarray}
S^{(NR)}=-\frac{\sqrt{\lambda_{NR}}}{4 \pi}\int d^2\sigma \mathcal{L}^{(NR)}_{P},
\end{eqnarray}
where the corresponding NR Lagrangian density is defined as\footnote{For the purpose of this paper, we restrict ourselves only to the NS-NS sector of the full supergravity solution. The reason behind this stems from the fact that the TSNC data corresponding to the RR sector is not yet settled down. Therefore, a priori, the NR limit (those have been obtained in Appendix \ref{appen A}) of the RR sector is not clear from the perspective of the geometric data in the NR sector. A complete understanding of these TSNC data would lead towards a supersymmetric generalization of the F-string Galilean algebra. However, as an interesting fact, we wish to point out that one can still calculate the number of ($ p ,q $) five branes (for the nonrelativistic theory) starting from the basic definition of Page charges as in a relativistic set up and thereby taking a $ c \rightarrow \infty $ limit of that (see sections \ref{2.2.4} and \ref{2.3.3}). In other words, without knowing the explicit TSNC data for the RR sector, one can still figure out the five brane configuration in a nonrelativistic set up following simple scaling arguments along the holographic ($ \eta $) axis.}
\begin{eqnarray}
\label{e2.28}
\mathcal{L}^{(NR)}_{P} =\sqrt{- \gamma}\gamma^{\alpha \beta} e_{\mu}^a e_{\nu}^b \partial_{\alpha}X^{\mu}\partial_{\beta}X^{\nu}\delta_{ab}+\eta_{AB}(\tau_{\mu}^A\pi_{\nu}^B - \tau_{\nu}^A\pi_{\mu}^B)\dot{X}^{\mu}X'^{\nu}\nonumber\\
+\delta_{ab}(e_{\mu}^a \pi_{\nu}^b -e_{\nu}^a \pi_{\mu}^b)\dot{X}^{\mu}X'^{\nu}+\zeta \varepsilon^{\alpha \beta}e_{\alpha}^+ \tau_{\mu}^+ \partial_{\beta}X^{\mu}+\bar{\zeta}\varepsilon^{\alpha \beta}e_{\alpha}^- \tau_{\mu}^-\partial_{\beta}X^{\mu}.
\end{eqnarray}

Here, we introduce $ \zeta $ and $ \bar{\zeta} $ as world-sheet scalars together with $ e_{\alpha}^{\pm}= e_{\alpha}^0 \pm e_{\alpha}^1$ and $ \tau_{\mu}^{\pm}=\tau_{\mu}^0 \pm \tau_{\mu}^1 $. The above Lagrangian (\ref{e2.28}) can be further simplified by choosing the conformal gauge $ \det \gamma_{\alpha \beta}=-1 $ for the world-sheet vielbeins.

The resulting Lagrangian density turns out to be
\begin{eqnarray}
\label{e2.29}
-\mathcal{L}^{(NR)}_{P} &=&h_{ij}\partial_{\alpha}X^{i}\partial_{\beta}X^{j}\eta^{\alpha \beta}+\varepsilon^{\alpha \beta}B^{(NR)}_{ij}\partial_{\alpha}X^{i}\partial_{\beta}X^{j}-(\zeta -\bar{\zeta})(\tau_{\hat{\mu}}^{0}\dot{X}^{\hat{\mu}}-\tau_{\hat{\mu}}^1 X'^{\hat{\mu}}),\\
B_{ij}^{(NR)}&=&\delta_{ab}e_{[i}^a\pi^b_{j]}~;~i,j=\chi , \xi
\end{eqnarray}
where, $ X^{\hat{\mu}} $ characterise the longitudinal directions together with the transverse metrics as
\begin{eqnarray}
h_{\chi \chi} =\frac{\pi \eta}{3\sqrt{2}}~;~h_{\xi \xi}=\frac{\pi \eta \chi^2}{3\sqrt{2}}.
\end{eqnarray}

\paragraph{Symmetries.} The Lagrangian (\ref{e2.29}) has a $ SO(1,1) $ boost symmetry generated by $ K_{01}$  associated with the longitudinal directions $X^0= t , X^1=\eta $
\begin{eqnarray}
\delta \tau_{\hat{\mu}}^{\alpha}=\Lambda_{\hat{\mu}}^{\hat{\nu}}\tau^{\alpha}_{\hat{\nu}}~;~\delta X^{\hat{\mu}}=\Lambda^{\hat{\mu}}_{\hat{\nu}}X^{\hat{\nu}},
\end{eqnarray} 
where we identify the matrices $ \Lambda \in SO(1,1) $.

On the other hand, associated with the transverse $X^{i} (=\chi , \xi)$ directions, one can imagine a $ SO(2) \subset SU(2)_R$ rotational invariance of the form
\begin{eqnarray}
\delta X^i = \lambda^i_j X^j ,
\end{eqnarray}
where the matrices are identified as $ \lambda \in SO(2) $. These rotations can be associated with angular momentum generators of the form $ L_{ij} $. 

On top of these, the sigma model possesses translation symmetries $ \delta \xi =\xi_0 $ and $ \delta t=t_0 $. These two directions are what we identify as the isometries associated with the non-Lorentzian target space. 

The translation along $ \xi $ is generated by the transverse translation generators $ P_{\xi}=e_{\xi}^3P_3 $ and the time translation is generated by nonrelativistic world-sheet Hamiltonian $ H_t = \tau_t^0 H_0 $. A closer look further reveals that, under string Galilean boost \cite{Bidussi:2021ujm} both the variations $ \delta h_{ij}=0=\delta B^{(NR)}_{ij} $ which stems from the fact that the longitudinal one form $ \tau_i^A =0 $.
\subsubsection{Laplace equation and the rank function}
The Laplace equation in the electrostatic description \cite{Legramandi:2021uds} is given by
\begin{eqnarray}
\partial^2_{\sigma}\hat{V}+\partial^2_{\eta}\hat{V}=0.
\end{eqnarray}

The associated rank function turns out to be \cite{Legramandi:2021uds}
\begin{eqnarray}
\label{eee2.35}
\mathcal{R}(\eta)=\partial_{\sigma}\hat{V}|_{\sigma =0}=c^2\tilde{\mathcal{R}}(\tilde{\eta}),
\end{eqnarray}
where, the rank function in the NR limit is identified as
\begin{eqnarray}
\label{2.34}
\tilde{\mathcal{R}}(\tilde{\eta})= \begin{cases}
      \frac{\tilde{\eta}N_c}{2}  &  0 \leq \tilde{\eta} \leq \frac{(P-1)}{c^2}\\
      \frac{N_c}{2} (P-1) (\frac{P}{c^2}-\tilde{\eta})&\frac{(P-1)}{c^2}\leq \tilde{\eta} \leq \frac{P}{c^2}.
    \end{cases} 
\end{eqnarray}

In the NR limit, one defines a modified boundary conditions for the rank function
\begin{eqnarray}
\tilde{\mathcal{R}}(\tilde{\eta}_{min}= 0)=0=\tilde{\mathcal{R}}\left(\tilde{\eta}_{max}=\frac{P}{c^2} \right), 
\end{eqnarray}
where $ \frac{P}{c^2} $ is the new location of the conducting plane that was initially at a location $\eta= P $.
\subsubsection{Brane set up and Page charges}
\label{2.2.4}
We now discuss the effect of NR scaling on NS5-D5-D7 brane configuration. In order to explore the brane set up, it is customary first to estimate the Page charges in the NR limit.

Let us first estimate the Page charge associated to NS5 branes
\begin{eqnarray}
Q_{NS5}=\frac{1}{4\pi^2}\int H_3 ~;~ H_3 = \partial_{\eta}f_4 d\eta \wedge \text{Vol}(S^2).
\end{eqnarray}

Considering the NR scaling (\ref{e2.18}), this finally leads to
\begin{eqnarray}
\label{ee2.39}
\tilde{Q}_{NS5}&=&\frac{1}{4\pi^2 c^2}\int_0^{P/c^2}d\tilde{\eta}\partial_{\tilde{\eta}}f_4(\sigma = \pm \infty , \tilde{\eta})\int_0^{c\pi}\tilde{\chi}d\tilde{\chi}\int_0^{2\pi}d\tilde{\xi}\nonumber\\
&=&\frac{\pi}{4}(f_4(\sigma = \pm \infty , \tilde{\eta}=P/c^2)-f_4(\sigma = \pm \infty , \tilde{\eta}=0)).
\end{eqnarray}

Estimating the metric function $ f_4 $ both at $ \sigma = \pm \infty $, finally leads to
\begin{eqnarray}
\label{e2.38}
\tilde{Q}_{NS5} = P.
\end{eqnarray}

\begin{figure}
\includegraphics[scale=.75]{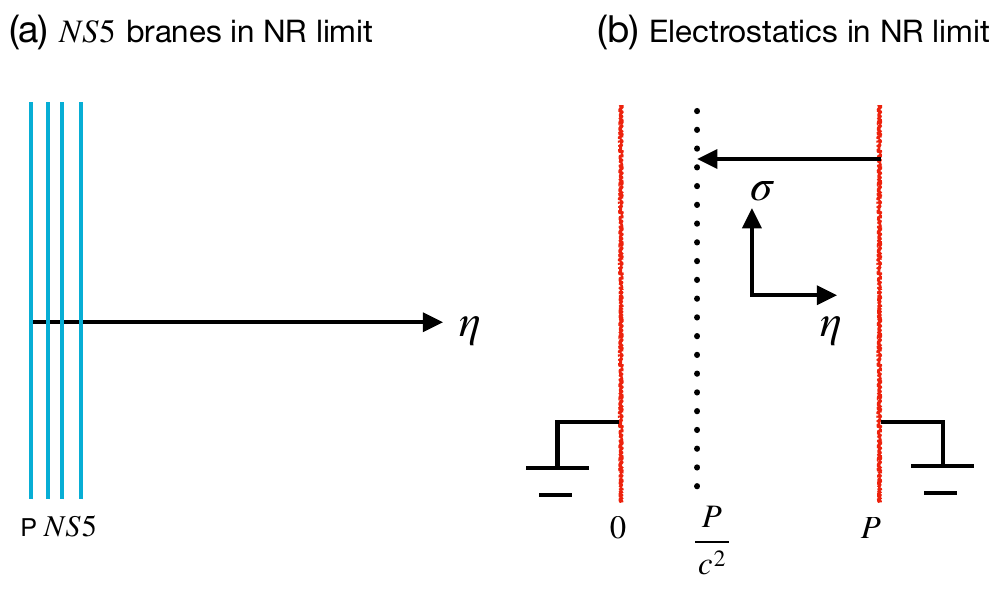}
  \caption{(a) NS5 brane configuration in the TSNC limit. $P$ NS5 branes are localised near the origin $\eta \sim 0$ of the holographic axis as result of TSNC scaling (\ref{e2.18}). (b) The conducting plates of electrostatic description \cite{Legramandi:2021uds} are now closely spaced as a result of TSNC scaling (\ref{e2.18}).} \label{hw2}
\end{figure}

Modulo an overall scaling ($ \sim \frac{\pi^2}{4} $) which may be absorbed in the definition of $ \tilde{Q}_{NS5} $, the above relation (\ref{e2.38}) simply ensures the conservation of NS5 brane charge in the TSNC limit. However, the location of these NS5 branes are now rescaled/shifted by a factor of $ 1/c^2 $ which we identify as an artefact of the TSNC scaling (\ref{e2.18}). 

To summarise, therefore in the strict TSNC ($ c \rightarrow \infty $) limit, all these $ P $ NS5 branes are eventually put on top of each other near the origin ($ \eta \sim 0 $) of the holographic axis causing a singularity there (see Fig.\ref{hw2}). This effect is precisely reflected as a singularity in the one form $ \tau_{\eta}^1 $ (\ref{2.23}) in the limit, $ \eta \rightarrow 0 $. 

For D7 branes, we have the following expression for the Page charge (following a proper rescaling) in the relativistic set up \cite{Legramandi:2021uds}
\begin{eqnarray}
Q_{D7} = \mathcal{R}'(0)-\mathcal{R}'(P)= N_c P.
\end{eqnarray}

Using (\ref{eee2.35}) (and following a suitable rescaling), one can show that in the TSNC limit, $ \mathcal{R}'(\eta)=\tilde{\mathcal{R}}'(\tilde{\eta}) $ which therefore leads towards the conservation of D7 brane charge as follows
\begin{eqnarray}
\label{2.40}
 \tilde{Q}_{D7}=\tilde{\mathcal{R}}'(0)-\tilde{\mathcal{R}}'(P/c^2)  =N_c P.
\end{eqnarray}

Finally, we note down the D5 brane charge and its NR limit. In the type IIB description, for an interval $ \eta \in [k, k+1]$, one finds the D5 brane charge goes as \cite{Legramandi:2021uds}
\begin{eqnarray}
\label{e2.43}
Q_{D5}[k , k+1]=\frac{4}{\pi}\left( \mathcal{R}(\eta)-\mathcal{R}'(\eta)\left(\eta - k \right)\right) .
\end{eqnarray}

Using (\ref{2.34}), we finally note down the following scaling relation in the TSNC limit
\begin{eqnarray}
\label{e2.44}
Q_{D5}&=&\int_0^P \mathcal{R}(\eta)d\eta =\int_0^{P/c^2}\tilde{\mathcal{R}}(\tilde{\eta}) d \tilde{\eta}=  \frac{\tilde{Q}_{D5}}{2c^4},
\end{eqnarray}
where $ \tilde{Q}_{D5} =\frac{N_c}{2}  P (P-1)$ counts the number of color D5 branes in the nonrelativistic limit. One should think of (\ref{e2.44}) as a simple nonrelativistic scaling relation of the form $ Q_{D5} \rightarrow \frac{Q_{D5}}{c^4} $ where the total number of D5 branes is preserved in the nonrelativistic limit.

Let us try to understand the $ \frac{1}{c^4} $ factor sitting in front in (\ref{e2.44}). As a result of nonrelativistic scaling, the NS5 branes are now all shifted by a factor of $ \frac{\eta_i}{c^2} $ along the holographic ($ \eta $) axis which produces a factor of $ \frac{1}{c^2} $. On top of this, the quiver appears with a single \emph{kink} which corresponds to a (positive) slope (see Fig. \ref{hw1}b). It is this slope that produces an additional factor of $ \frac{1}{c^2} $ while integrating the rank function along the holographic axis.
\subsubsection{TSNC scaling at special points}
\label{2.2.5}
As a final remark, we wish to explore the properties of the metric functions and their NR scaling in the large $ \sigma = \Lambda \rightarrow \pm \infty $ limit. Using the potential function (\ref{2.14}), below we enumerate the metric functions in the large $ \sigma $ limit. 
\begin{eqnarray}
f_1 (\sigma \sim \Lambda , \eta \sim 0)&=&\pm \frac{3 \pi  \Lambda }{2}+\frac{9 P}{4}+\mathcal{O}(1/\Lambda ),\\
f_2 (\sigma \sim \Lambda , \eta \sim 0)&=&\frac{\pi ^2 \eta ^2}{2 P}+\mathcal{O}(1/\Lambda ),\\
f_3 (\sigma \sim \Lambda , \eta \sim 0)&=&\frac{\pi ^2}{2 P}+\mathcal{O}(1/\Lambda ),\\
f_4 (\sigma \sim \Lambda , \eta \sim 0)&=&\frac{\pi ^3 \eta ^3}{3 P^2}+\mathcal{O}(1/\Lambda ).
\end{eqnarray}

To obtain the corresponding TSNC data, we propose the following NR scaling
\begin{eqnarray}
\label{ET2.51}
t =c \tilde{t}~;~\eta = c \tilde{\eta}~;~\xi = \tilde{\xi}~;~\chi =\frac{\tilde{\chi}}{c} .
\end{eqnarray}

This leads to the following expressions for the longitudinal as well as transverse vielbeins in the large $c \rightarrow \infty$ limit.
\begin{eqnarray}
\tau_t^0  &=&\sqrt{| \pm \frac{3 \pi  \Lambda }{2}+\frac{9 P}{4}|}~;~\tau_{\eta}^1 = \frac{\pi}{\sqrt{2P}},\\
e_{\chi}^2 &=& \frac{\pi \eta}{\sqrt{2P}}~;~e_{\xi}^3 = \frac{\pi \eta \chi}{\sqrt{2P}}.
\end{eqnarray}

The NS-NS two from, on the other hand, turns out to be
\begin{eqnarray}
B_2 = c B^{(NR)}_2 ~;~ B^{(NR)}_2=\frac{\pi^3 \eta^3 \chi}{3P^2}.
\end{eqnarray}

Clearly, as one can see, unlike in the previous example, all the TSNC data are non-singular in the limit $ \eta \sim 0 $. In other words, the singularity at the origin of the $ (\sigma , \eta) $ coordinate system is not visible from a large distance along the $ \sigma $ axis.
\subsection{$  +_{P,N_c}$ quivers}
\begin{figure}
\includegraphics[scale=.85]{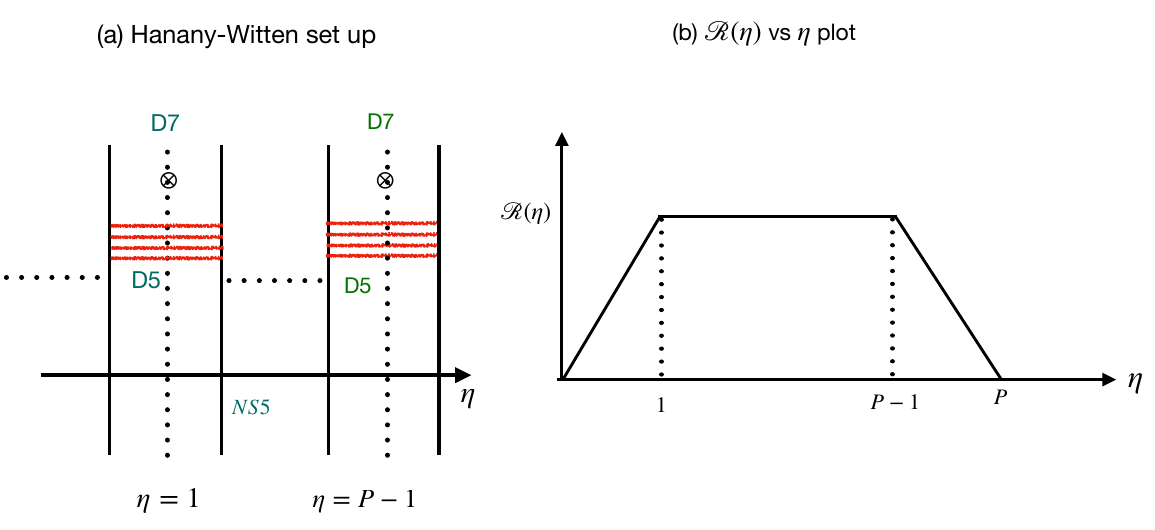}
  \caption{(a) NS5-D5-D7 brane intersections for $  +_{N_c,P}$ quivers. Flavour D7 branes are located at $\eta =1$ and $\eta =P-1$ along the holographic axis. (b) The corresponding rank function $\mathcal{R}(\eta)$ exhibits a plateau for $1 \leq \eta \leq P-1$.} \label{hw3}
\end{figure}

These quivers are of special interest because of their richer structure as compared to the previous one. The corresponding rank function ($ \mathcal{R}(\eta) $) is piecewise linear and possesses a plateau as a consequence of the location of flavour D7 branes at distinct locations along the holographic axis (see Fig.\ref{hw3}).

The rank function in this case reads as \cite{Legramandi:2021uds}, \cite{Roychowdhury:2021jqt}
\begin{eqnarray}
\label{e2.52}
\mathcal{R}(\eta)= \begin{cases}
      N_c \eta &  0 \leq \eta \leq 1\\
      N_c  &1\leq \eta \leq (P-1)\\
      N_c (P-\eta) &(P-1)\leq \eta \leq P,
    \end{cases} 
\end{eqnarray}
which corresponds to placing flavour D7 branes at $ \eta =1 $ and $ \eta = P-1 $ (see Fig.\ref{hw3}).

Clearly, the presence of flavour branes at distinct locations modifies the geometry. As a result, we categorise the background into three different regions which we list below. 

A careful analysis reveals that when expanded near $ \sigma \sim 0 $, the corresponding potential function $ \hat{V}(\sigma , \eta) $ reads as
\begin{eqnarray}
\frac{\hat{V}(\sigma , \eta)}{(\frac{N_c}{4 \pi})}\sim \eta  (6+4\log 2)-4 \eta  \log \left(\frac{\pi }{P}\right)-2\eta \log |1 - \eta^2 | -(1+ \eta^2 - \sigma^2) \log \left| \frac{\eta +1}{\eta -1}\right|,
\end{eqnarray}
which turns out to be regular \cite{Roychowdhury:2021jqt} across the location of the flavour D7 branes at $ \eta =1 $.

As mentioned above, we divide the entire range $ 0 \leq \eta \leq P $ into following three regions.
\paragraph{Region I.} This region corresponds to the range $ 0 \leq \eta < 1 $. The resulting metric functions read as
\begin{eqnarray}
f_1 (\sigma \sim 0, \eta <1)& \sim & \frac{3}{2} \pi  \sqrt{|-\frac{\eta ^2}{2}-3 \log \left(\frac{\pi }{P}\right)+3+\log 8|},\\
f_2 (\sigma \sim 0, \eta <1)& \sim & -\frac{\pi  \eta ^2 \left(-\eta ^2-6 \log \left(\frac{\pi }{P}\right)+6+\log (64)\right)^{3/2}}{3 \left(\sqrt{2} \left(\eta ^4+8 \eta ^2 \left(\log \left(\frac{\pi }{2 P}\right)-1\right)-4 \left(\log \left(\frac{\pi }{2 P}\right)-1\right)^2\right)\right)},\\
f_3 (\sigma \sim 0, \eta <1)& \sim & -\frac{3 \left(\pi  \sqrt{|-\frac{\eta ^2}{2}-3 \log \left(\frac{\pi }{P}\right)+3+\log 8|}\right)}{\eta ^2+6 \log \left(\frac{\pi }{P}\right)-6 (1+\log 2)}.
\end{eqnarray}
\paragraph{Region II.} This is the region which corresponds to $ 1- \delta \leq \eta \leq 1+ \delta $ where $ \delta $ being very small. The corresponding metric functions read as
\begin{eqnarray}
f_1 (\sigma \sim 0, \eta \sim 1)& \sim &\frac{3\sqrt{3}\pi}{\sqrt{2}} k_c \sqrt{\left|2 \log \left(\frac{\pi }{P}\right)-3\right|},\\
f_2 (\sigma \sim 0, \eta \sim 1)& \sim &\frac{1}{9}f_1 (\sigma \sim 0, \eta \sim 1),\\
f_3 (\sigma \sim 0, \eta \sim 1)& \sim &\frac{9\pi^2}{4}f^{-1}_1 (\sigma \sim 0, \eta \sim 1),
\end{eqnarray} 
with $ k_c = \frac{\delta}{\sigma} $ is kept fixed in the limit $ \delta \rightarrow 0 $.
\paragraph{Region III.} Finally, we consider the region $ 1+\delta \leq \eta \leq P $ which corresponds to metric functions of the form\footnote{See Appendix \ref{appendixA} for details. }
\begin{eqnarray}
 f_1(\sigma \sim 0, \eta > 1) &\sim & \frac{3\sqrt{3}}{2\sqrt{2}}  \pi  \sqrt{\frac{a_1(\eta)}{b_1(\eta)}},\\
 f_2(\sigma \sim 0, \eta > 1) &\sim & \frac{a_2(\eta)}{b_2(\eta)},\\
 f_3 (\sigma \sim 0, \eta > 1)&\sim& -\frac{a_3(\eta)}{b_3(\eta)}.
\end{eqnarray}
\subsubsection{Decoding the vielbeins}
We obtain TSNC data for two different regions of the spacetime namely: (i) $ 0 \leq \eta \leq 1 $ and (ii) $ 1< \eta \leq P $. The NR scaling is defined as 
\begin{eqnarray}
\label{2.65}
t =\tilde{t}~;~ \eta = c^2 \tilde{\eta} ~;~\xi = \tilde{\xi}~;~\chi =\frac{\tilde{\chi}}{c}.
\end{eqnarray}

For the case (i), the resulting TSNC data turns out to be
\begin{eqnarray}
\label{2.66}
\tau_t^0 =\sqrt{\frac{3 \pi \eta}{2\sqrt{2}}}~;~ \tau_{\eta}^1 = \sqrt{\frac{3 \pi}{\sqrt{2}\eta}}~;~e_{\chi}^2 =\sqrt{\frac{\pi \eta}{3\sqrt{2}}}~;~e_{\xi}^3 =\sqrt{\frac{\pi \eta}{3\sqrt{2}}} \chi,
\end{eqnarray}
where we remove tildes for simplicity.

On the other hand, for the case (ii), in the large $ c $ limit of the vielbeins one finds
\begin{eqnarray}
c E_t^0 &=& c \sqrt{\frac{3\sqrt{3}\pi \tilde{\eta}}{2\sqrt{2}}}\left( 1+\frac{1}{4\tilde{\eta}c^2} \mathcal{K}(c^2 \tilde{\eta})+ \cdots\right) =c \left( \tau_t^0 - \frac{1}{2c^2}\pi_t^1 \right),\\
\mathcal{K}(\eta)&=&\frac{1}{\log \frac{\eta +1}{\eta -1}}( 2\log (\eta^2 -1) -6  -4 \log 2 +4 \log (\frac{\pi}{P})).
\end{eqnarray} 

A straightforward comparison reveals the following longitudinal one forms
\begin{eqnarray}
\tau_t^0 &=& \sqrt{\frac{3\sqrt{3}\pi \tilde{\eta}}{2\sqrt{2}}} \left( 1+\frac{1}{4}\left( \log \left(\frac{c^4 \tilde{\eta} ^2}{4}\right)+2 \log \left(\frac{\pi }{P}\right)-3\right) \right),\\
\pi_t^1 &=&0.
\end{eqnarray}

On a similar note, one finds
\begin{eqnarray}
\label{2.70}
c E_{\tilde{\eta}}^1= c\left(\frac{3}{2} \right)^{1/4} \sqrt{\frac{\pi}{\tilde{\eta}}}\left(1-\frac{1}{4 \tilde{\eta} c^2}\mathcal{K}(c^2 \tilde{\eta})+ \cdots\right)=c (\tau_{\tilde{\eta}}^1-\frac{1}{2c^2}\pi_{\tilde{\eta}}^0 ),
\end{eqnarray}

From (\ref{2.70}), it trivially follows that
\begin{eqnarray}
\label{2.72}
\tau_{\tilde{\eta}}^1 &=&\left(\frac{3}{2} \right)^{1/4} \sqrt{\frac{\pi}{\tilde{\eta}}}\left( 1-\frac{1}{4}\left( \log \left(\frac{c^4 \tilde{\eta} ^2}{4}\right)+2 \log \left(\frac{\pi }{P}\right)-3\right) \right),\\
\pi_{\tilde{\eta}}^0 &=&0.
\end{eqnarray}

Before proceeding further, let us first explore the behaviour of TSNC fields near $ \eta \sim 0 $. Like in the previous example, we notice that $ \tau_{\eta}^1 $ (\ref{2.66}) becomes singular as one approaches the origin of the holographic axis. This singularity, like before, is an artefact of the collapsing NS5 branes as a result of the NR scaling (\ref{2.65}).

In the strict NR limit, the range $ \eta \in [0, P] $ becomes singular as result of the $ 1/c^2 $ scaling. This corresponds to the fact that the NS5 branes those were in the region $ \eta > 1 $ are now all collapsed at the singularity $ \eta \sim 0 $. This is reflected as a singularity of the vielbein (\ref{2.72}) in the limit $ \tilde{\eta} \rightarrow 0$.

Finally, we note down transverse vielbeins those read as
\begin{eqnarray}
e_{\tilde{\chi}}^2 = \sqrt{\frac{\pi \tilde{\eta}}{2}}\left(\frac{3}{2}\right)^{1/4}~;~e_{\tilde{\xi}}^3 =\sqrt{\frac{\pi \tilde{\eta}}{2}}\left(\frac{3}{2}\right)^{1/4}\tilde{\chi}.
\end{eqnarray}

\begin{figure}
\includegraphics[scale=.7]{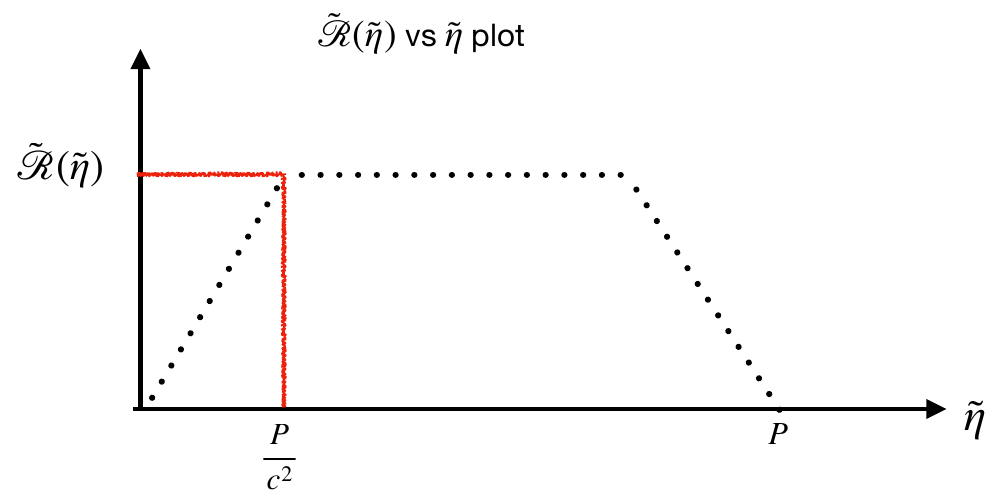}
  \caption{We plot the rank function against the rescaled holographic axis ($\tilde{\eta}$) in the TSNC limit of $\mathcal{N}=1$ quivers. The red lines show the modified rank function in the large $c$ limit while the dotted lines correspond to the rank function in the relativistic scenario. The rank function vanishes sharply at the end points while exhibiting a plateau in the intermediate regime.} \label{hw4}
\end{figure}
\subsubsection{Laplace equation and the rank function}
The associated charge density/ the rank function is defined as \cite{Legramandi:2021uds}
\begin{eqnarray}
\mathcal{R}(\eta)=\lim_{\epsilon \rightarrow 0}(\partial_{\sigma}\hat{V}(\sigma  =+\epsilon , \eta) - \partial_{\sigma}\hat{V}(\sigma  =-\epsilon , \eta)).
\end{eqnarray}

A straightforward computation reveals
\begin{eqnarray}
\frac{\mathcal{R}(\eta)}{\frac{N_c}{\pi}}~~~~~~~~~~~~~~~~~~~~~~~~~~~~~~~~~~~~~~~~~~~~~~~~~~~~~~~
~~~~~~~~~~~~~~~~~~~~~~~~~~~~~~~~~~~~~~~~\nonumber\\
=\text{Re}\left(  i \left((\eta -1) \log \left(-\frac{i \pi  (\eta -1)}{P}\right)-(\eta +1) \log \left(-\frac{i \pi  (\eta +1)}{P}\right)+2+\log 4\right)\right) 
\end{eqnarray}
which by virtue of the TSNC scaling (\ref{2.65}) reveals the NR charge density as
\begin{eqnarray}
\tilde{\mathcal{R}}(\tilde{\eta})=4 \pi N_c \arg \left(-\frac{i \tilde{\eta} }{P}\right).
\end{eqnarray}

Clearly, the charge density vanishes at the origin namely $ \tilde{R}(\tilde{\eta}_{min}=0)=0 $. On the other hand, for the other end point, we find $  \tilde{R}(\tilde{\eta}_{max})= 4 \pi N_c \arg \left(-\frac{i}{c^2}\right) $ which also vanishes in the strict large $ c \rightarrow \infty $ limit. Therefore, to summarise, both the boundary conditions are preserved in the TSNC limit of $\mathcal{N}=1$ quivers. 

On the other hand, in the intermediate region $ 0 < \tilde{\eta}< \frac{P}{c^2} $, the rank function $\tilde{\mathcal{R}}(\tilde{\eta})$ exhibits a plateau region as in the relativistic scenario (see Fig.\ref{hw4}). Combining these two features together, we propose an alternative expression for the rank function in the TSNC limit which reads as
\begin{eqnarray}
\tilde{\mathcal{R}}(\tilde{\eta})= \begin{cases}
      N_c  &  \frac{1}{c^2} \leq \tilde{\eta} \leq \frac{(P-1)}{c^2}\\
      N_c (\frac{P}{c^2}-\tilde{\eta}) &\tilde{\eta}=\frac{P}{c^2}.
    \end{cases} 
\end{eqnarray}

From Fig.\ref{hw4}, it is quite evident that the area under the curve is much less as compared to its relativistic counterpart. This clearly indicates a lower charge density in the TSNC limit of the $\mathcal{N}=1$ quivers.

We probe more on this in the next section, where we compute Page charges and explore the associated NS5-D5-D7 brane set up in the TSNC limit.
\subsubsection{Brane set up and Page charges}
\label{2.3.3}
We begin by computing the Page charge corresponding to NS5 branes. To begin with, we first note down the potential function in the large $ \sigma \rightarrow \infty $ limit
\begin{eqnarray}
\label{2.78}
\hat{V}(\sigma \rightarrow \infty , \eta) \simeq -\frac{i N_c P}{4} -\left( \frac{\eta N_c \log \left(\frac{\pi  \sigma }{2 P}\right)}{\pi }+\frac{i N_c \sigma  \left(\log \left(\frac{\pi  \sigma }{2 P}\right)-1\right)}{\pi }\right) \nonumber\\
\frac{1}{P^2}(-\frac{1}{36} i \pi  \left(3 \eta ^2+1\right)N_c \sigma -\frac{1}{36} \pi  \eta  \left(\eta ^2+1\right) N_c+\frac{1}{12} \pi  \eta N_c \sigma ^2+\frac{1}{36} i \pi  N_c \sigma ^3)
\end{eqnarray}

Using (\ref{2.78}) and considering the real part, one finally obtains
\begin{eqnarray}
f_4 (\sigma = \pm \infty , \tilde{\eta})|_{c \rightarrow \infty} = \frac{2}{3} \pi  c^2\tilde{\eta},
\end{eqnarray}
which by means of (\ref{ee2.39}) reveals the NS5 brane Page charge in the TSNC limit as
\begin{eqnarray}
\tilde {Q}_{NS5} =P.
\end{eqnarray}

Clearly, following a suitable rescaling, like in the previous example, one precisely conserves the number of ($ =P $) NS5 branes in the TSNC limit. As we emphasised before, in the large $ c \rightarrow \infty $ limit, all these $ P $ NS5 branes are pushed together near the origin of the holographic axis (see Fig. \ref{hw2}(a)) creating a metric singularity there.

On a similar note, the Page charge corresponding to D7 branes turns out to be
\begin{eqnarray}
\tilde{Q}_{D7}=(\tilde{\mathcal{R}}'(0)-\tilde{\mathcal{R}}'(\tilde{\eta}_{max}))=N_c ,
\end{eqnarray}
which corresponds to the $ N_c $ flavour D7 branes localised at $ \tilde{\eta}=\tilde{\eta}_{max}=\frac{P}{c^2} $ (see Fig.\ref{hw4}).

Finally, we note down the number of color D5 branes in the TSNC limit 
\begin{eqnarray}
Q_{D5}=\frac{\tilde{Q}_{D5}}{c^2},
\end{eqnarray}
where $ \tilde{Q}_{D5} =N_c (P -1)$ is the number of color D5 branes in the nonrelativistic limit. The above result simply follows from the fact that there are $ (P-1) $ $ N_c $ color nodes whose relative spacing is now rescaled by a factor of $ \frac{1}{c^2} $ as an artefact of nonrelativistic scaling of the original positions ($ \eta_i $) of NS5 branes along the holographic axis. 

Notice that, unlike the previous example of single kink quivers, here we have a factor of $ \frac{1}{c^2} $ floating around. This is simply because of the fact that the quiver is \emph{flat}. In other words, it is with zero slope and therefore does not produce an additional factor of $ \frac{1}{c^2} $.
\subsubsection{TSNC scaling at special points}
We complete our discussion on TSNC limit by exploring the metric functions in the asymptotic limits ($ \sigma = \Lambda \rightarrow \pm \infty $) and thereby decoding the vielbeins in those limits.

Below, we enumerate metric functions those directly follow from (\ref{2.78})
\begin{eqnarray}
\label{2.85}
f_1 (\sigma \sim \Lambda , \eta \sim 0) &=&\frac{9P}{\sqrt{2}}~;~f_2 (\sigma \sim \Lambda , \eta \sim 0) =2\sqrt{2}P,\\
f_3 (\sigma \sim \Lambda , \eta \sim 0) &\sim & \mathcal{O}(P/\Lambda^2)~;~f_4 (\sigma \sim \Lambda , \eta \sim 0) \sim 0.
\label{2.86}
\end{eqnarray}

Based on the above information (\ref{2.85})-(\ref{2.86}), we propose TSNC scaling as
\begin{eqnarray}
t =c \tilde{t}~;~\eta = c \tilde{\eta}~;~\xi = \tilde{\xi}~;~\chi =\tilde{\chi},
\end{eqnarray}
which yield the following TSNC data
\begin{eqnarray}
\tau_t^0 = \frac{3P}{2^{\frac{1}{4}}}~;~\tau_{\eta}^1 \sim \mathcal{O}(1/\Lambda^2)~;~e_{\chi}^2 = \sqrt{2\sqrt{2}P}~;~e_{\xi}^3=\sqrt{2\sqrt{2}P}\sin\chi ~;~B^{(NR)}_2 \sim 0.
\end{eqnarray}
\section{T-duality}
Our purpose here is to discuss the effects of applying T-duality on the NR sigma model (\ref{e2.29}) and use the above as tool to compute various field theory observables in the TSNC limit of $ \mathcal{N}=1 $ quivers in 5d. Below, we present a general algorithm to obtain T-dual Lagrangian using the canonical framework of \cite{Alvarez:1994wj}.

We begin by considering the TSNC sigma model (\ref{e2.29}) in its most generic form
\begin{eqnarray}
\label{3.1}
\mathcal{L}^{(NR)}_P=h_{\hat{i}\hat{i}}(\dot{X}^{\hat{i}})^2 + 2 h_{\hat{i}m}\dot{X}^{\hat{i}}\dot{X}^m +h_{mn}\dot{X}^m \dot{X}^n -h_{\hat{i}\hat{i}}(X'^{\hat{i}})^{2}
-2h_{\hat{i}m}X'^{\hat{i}}X'^m \nonumber\\
-h_{mn}X'^m X'^n -2B^{(NR)}_{\hat{i}m}(X'^m \dot{X}^{\hat{i}}-X'^{\hat{i}}\dot{X}^m)+(\zeta -\bar{\zeta})(\tau_{\hat{\mu}}^{0}\dot{X}^{\hat{\mu}}-\tau_{\hat{\mu}}^1 X'^{\hat{\mu}}).
\end{eqnarray}

Here, $ X^i = \lbrace X^{\hat{i}},X^m\rbrace $ stand for the transverse coordinates associated with TSNC manifold. On the other hand, $ X^{\hat{\mu}}(\hat{\mu}=0,1) $ represent longitudinal directions.

In particular, $ X^{\hat{i}}(= \xi) $ is identified as the \emph{isometric} circular direction (associated to TSNC manifold) along which the T-duality is applied. The corresponding generating function is defined as \cite{Alvarez:1994wj}
\begin{eqnarray}
G = \frac{1}{2}\int d \sigma^1 (X'^{\hat{i}}\tilde{X}^{\tilde{i}}-X^{\hat{i}}\tilde{X}'^{\tilde{i}}),
\end{eqnarray}
where $ \tilde{X}^{\tilde{i}} $ is introduced as the dual coordinate that is accompanied by a dual momentum 
\begin{eqnarray}
 \tilde{p}_{\tilde{i}}=-\frac{\partial G}{\partial \tilde{X}^{\tilde{i}}}=-X'^{\hat{i}}~;~p_{\hat{i}}=\frac{\partial G}{\partial X^{\hat{i}}}=-\tilde{X}'^{\tilde{i}}.
\end{eqnarray}

Below, we enumerate the canonically conjugate momenta those readily follow from (\ref{3.1})
\begin{eqnarray}
\label{3.4}
p_{\hat{\mu}}&=&(\zeta -\bar{\zeta})\tau_{\hat{\mu}}^{0},\\
p_m &=&2( h_{\hat{i}m}\dot{X}^{\hat{i}}+h_{mn}\dot{X}^n)+2B^{(NR)}_{\hat{i}m}X'^{\hat{i}},\\
p_{\hat{i}}&=&2(h_{\hat{i}\hat{i}}\dot{X}^{\hat{i}}+h_{\hat{i}m}\dot{X}^m)-2 B^{(NR)}_{\hat{i}m}X'^m.
\label{3.6}
\end{eqnarray}

The above relations (\ref{3.4})-(\ref{3.6}) can be inverted to obtain velocities in terms of momenta
\begin{eqnarray}
\dot{X}^m &=&h^{mn}(\frac{1}{2}p_n - B^{(NR)}_{\hat{i}n}X'^{\hat{i}}),\\
\dot{X}^{\hat{i}}&=&\frac{1}{h_{\hat{i}\hat{i}}}(\frac{1}{2}p_{\hat{i}}+B^{(NR)}_{\hat{i}m}X'^{m}),
\end{eqnarray}
where we set, $ h_{m\hat{i}}=0 $ which is compatible with TSNC backgrounds that we consider here.

This finally yields the T-dual Hamiltonian of the following form 
\begin{eqnarray}
\label{3.9}
\tilde{\mathcal{H}}=\frac{1}{4h_{\hat{i}\hat{i}}}(\tilde{X}'^{\tilde{i}})^2+\frac{h^{mn}}{4}p_m p_n +g_{mn}X'^m X'^n +g_{\hat{i}\hat{i}}(\tilde{p}_{\tilde{i}})^2\nonumber\\
+h^{mn} B^{(NR)}_{\hat{i}n}p_m\tilde{p}_{\tilde{i}}-\frac{B^{(NR)}_{\hat{i}m}}{h_{\hat{i}\hat{i}}}X'^m \tilde{X}'^{\tilde{i}}+(\zeta - \bar{\zeta})\tau_{\hat{\mu}}^1 X'^{\hat{\mu}},
\end{eqnarray}
where we define the following entities as
\begin{eqnarray}
g_{mn}&=&h_{mn}+\frac{1}{h_{\hat{i}\hat{i}}}B^{(NR)}_{m\hat{i}}B^{(NR)}_{n\hat{i}},\\
g_{\hat{i}\hat{i}}&=&h_{\hat{i}\hat{i}}+h^{mn}B^{(NR)}_{m\hat{i}}B^{(NR)}_{n\hat{i}}.
\end{eqnarray}

Velocities those readily follow from the dual Hamiltonian (\ref{3.9}) could be expressed as
\begin{eqnarray}
\label{3.12}
\dot{\tilde{X}}^{\tilde{i}}=2g_{\hat{i}\hat{i}}\tilde{p}_{\tilde{i}}+h^{mn}p_m B^{(NR)}_{\hat{i}n}~;~\dot{X}^m = \frac{h^{mn}}{2}p_n + h^{mn} B^{(NR)}_{\hat{i}n}\tilde{p}_{\tilde{i}}.
\end{eqnarray}

The above relation (\ref{3.12}) can be inverted to express momenta in terms of velocities as
\begin{eqnarray}
\label{3.13}
\tilde{p}_{\tilde{i}}&=&\frac{1}{h_{\hat{i}\hat{i}}}(\frac{\dot{\tilde{X}}^{\tilde{i}}}{2}+\dot{X}^m B^{(NR)}_{m\hat{i}}),\\
p_m &=&2g_{mn}\dot{X}^n + \frac{B^{(NR)}_{m\hat{i}}}{h_{\hat{i}\hat{i}}}\dot{\tilde{X}}^{\tilde{i}}.
\label{3.14}
\end{eqnarray}

The T-dual Lagrangian is defined as
\begin{eqnarray}
\tilde{\mathcal{L}}=\tilde{p}_{\tilde{i}}\dot{\tilde{X}}^{\tilde{i}}+p_m \dot{X}^m + p_{\hat{\mu}}\dot{X}^{\hat{\mu}}-\tilde{\mathcal{H}},
\end{eqnarray}
which by means of the above set (\ref{3.13})-(\ref{3.14}) of data reveals
\begin{eqnarray}
\tilde{\mathcal{L}}=\tilde{g}_{\tilde{i}\tilde{i}}\eta^{\alpha \beta}\partial_{\alpha}\tilde{X}^{\tilde{i}}\partial_{\beta}\tilde{X}^{\tilde{i}}+\tilde{g}_{mn}\eta^{\alpha \beta}\partial_{\alpha}X^m \partial_{\beta}X^n \nonumber\\ + \tilde{g}_{m \tilde{i}}\eta^{\alpha \beta}\partial_{\alpha}X^m \partial_{\beta}\tilde{X}^{\tilde{i}}
+(\zeta -\bar{\zeta})(\tau_{\hat{\mu}}^{0}\dot{X}^{\hat{\mu}}-\tau_{\hat{\mu}}^1 X'^{\hat{\mu}}).
\end{eqnarray}

Below, we enumerate T-duality rules for the metric as well as NS fluxes
\begin{eqnarray}
\label{3.17}
\tilde{g}_{\tilde{i}\tilde{i}}=\frac{1}{4 h_{\hat{i}\hat{i}}}~;~\tilde{g}_{mn}=g_{mn}~;~\tilde{g}_{m\tilde{i}}=\frac{B^{(NR)}_{ m\hat{i}}}{h_{\hat{i}\hat{i}}}~;~\tilde{B}^{NR}_{\tilde{i}m}=0,
\end{eqnarray}
where the vanishing of the NS-NS fluxes in the T-dual picture has its root in the choice of the metric component $ h_{m\hat{i}}=0 $ that we had mentioned earlier.

Considering one specific example, below we decode these T-duality rules explicitly. We take the example of $ T_{N_c , P} $ quivers in their TSNC limit. The compact isometry is identified as $ \xi $ whose dual is denoted as $ \tilde{\xi} $. This yields the following T-dual TSNC (transverse) metric components those readily follow from (\ref{3.17})
\begin{eqnarray}
\tilde{g}_{\tilde{\xi}\tilde{\xi}}=\frac{3\sqrt{2}}{4 \pi \eta \chi^2}~;~\tilde{g}_{\chi \chi}=\frac{\pi \eta}{3\sqrt{2}}\left(1+\frac{9 \pi^4 \eta^4}{2 P^4 \log^2 2} \right)~;~\tilde{g}_{\chi \tilde{\xi}}= \frac{3 \pi^2}{\sqrt{2}P^2\log 2}\frac{\eta^2}{\chi}.
\end{eqnarray}

Clearly, some of these T-dual metric components diverges in the limit $ \eta \rightarrow 0 $ which is an artefact of the collapsing $ P $ NS5 branes those were discussed previously. As a final note, we find that the longitudinal vielbeins ($ \tau^{\alpha}_{\hat{\mu}} $) do not transform under T-duality.
\section{QFT observables in the TSNC limit}
\subsection{Central charge}
The general idea of this section is to consider the holographic central charge of the relativistic description \cite{Legramandi:2021uds} and thereby taking its TSNC limit. The resulting entity we claim to be the holographic central charge in the TSNC limit of $ \mathcal{N}=1 $ quiver.

Central charge in the relativistic description is found to be \cite{Legramandi:2021uds}
\begin{eqnarray}
\hat{c}_{hol} =\frac{2}{3 \pi^5} \int_0^P d\eta \int_{-\infty}^{+\infty} d\sigma \sigma^3 \partial_{\sigma}V \partial^2_{\eta}V.
\end{eqnarray}

Let us consider the potential function corresponding to $ T_{N_c ,P} $ quivers
\begin{eqnarray}
\label{4.2}
\hat{V}(\sigma , \eta)=\frac{P^3 N_c}{2 \pi ^3}\text{Re}\left[\text{Li}_3\left (-e^ {-\frac{\pi  (\sigma +i+i \eta )}{P}}\right)-\text{Li}_3\left(-e^ {-\frac{\pi  (\sigma -i+i \eta )}{P}}\right)\right],
\end{eqnarray}
where we define $ \hat{V}=\sigma V $.

Using (\ref{4.2}), a straightforward computation in the large $ P $ limit reveals
\begin{eqnarray}
\hat{c}_{hol} = \frac{N^2_c P^2 \Lambda}{36 \pi^5} \int_0^P  d \eta +\mathcal{O}(1/P),
\end{eqnarray}
where we fix the limits of the $ \sigma $ integral as $ [ - \Lambda , \Lambda ] $ where $ \Lambda $ being large.

Following our previous discussion, the TSNC limit is realized by setting $ \eta = c^2 \tilde{\eta} $ which finally yields the central charge in the nonrelativistic limit as
\begin{eqnarray}
\hat{c}_{hol}=\frac{N^2_c P^3 \Lambda}{36\pi^5 c^4}=\frac{\hat{c}_{nr}}{c^4}.
\end{eqnarray}

Clearly, the degrees of freedom associated to $ \mathcal{N}=1 $ SCFTs are reduced down significantly as one approaches the corresponding TSNC limit. This can also be interpreted using our previous results on Page charges associated to D5 branes which shows that the color degrees of freedom are reduced down to zero in the strict large $ c(\rightarrow \infty) $ limit.

A similar calculation for $ +_{P,N_c} $ quiver reveals
\begin{eqnarray}
\hat{c}_{hol} = \frac{N^2_c P \Lambda}{3\pi^6} \int_0^P  d \eta +\mathcal{O}(1/P),
\end{eqnarray}
which upon taking TSNC limit yields
\begin{eqnarray}
\hat{c}_{hol}=\frac{N^2_c P^2 \Lambda}{3 \pi^6 c^2} = \frac{\hat{c}_{nr}}{c^2}.
\end{eqnarray}
\subsection{Couplings}
Technically speaking, the coupling constant in the nonrelativistic counterpart of $ \mathcal{N}=1 $ quivers can be studied through color D5 brane probes in the bulk supergravity solution. These coupling constants are fixed by the VEV of the scalars $ \langle \Phi_i \rangle $ that corresponds to the location of the $ i $th NS5 brane along the holographic axis. 

These scalars $ \Phi_i $ form an important component of the tensor multiplet living on the world-volume of NS5 branes. One could imagine a typical coupling between the elements of the tensor multiplet and the elements of the vector mutiplet living on the world-volume of color D5 branes that gives rise to the Lagrangian of the form $ L \sim (\Phi_{i+1}-\Phi_i)F^2_{mn}+\cdots $.

Typically the QFT Lagrangian in 5d could be recast in the form
\begin{eqnarray}
\label{4.7}
S_{QFT}\sim \langle \Phi_{i+1}-\Phi_{i}\rangle \int d^5x F^2_{mn} \sim \frac{1}{g^2_{QFT}}\int d^5 x F^2_{mn},
\end{eqnarray}
which shows that the coupling constant goes inversely with the relative separation between NS5 branes. Nievely, one should therefore expect that the coupling constant in the nonrelativistic limit must grow large as the the NS5 branes are closely spaced.  

To see this explicitly, we consider the probe DBI action of the following form
\begin{eqnarray}
S_{DBI}=T_{D5}\int d^6 x \sqrt{-\det [g + 2 \pi \alpha' F]},
\end{eqnarray}
where we switch off the dilaton for the present computation and turn on world-volume gauge fields namely, $ F= F_{tx} $. 

The D5 brane is considered to be extended along the four Minkowski directions of the target space as well as along the holographic $ \eta $ direction. While it is considered to be located at a fixed position along the $ \sigma =0 $ and the radial ($ r=r_c $) direction of $ AdS_6 $.

The target space that we choose to work with turns out to be of the following form
\begin{eqnarray}
ds^2 = f_1(\eta) \frac{4 r^2_c}{L^2}dx^2_{1,4} +f_3(\eta)d\eta^2 ,
\end{eqnarray}
where the remaining directions are switched off.

The resulting DBI action takes the following form
\begin{eqnarray}
S_{DBI}=T_{D5}\left(\frac{4r^2_c}{L^2} \right)^{5/2}  \int d\eta d^5x \sqrt{f_3}f_1^{5/2}\sqrt{1+\frac{\pi^2 \alpha'^2 L^4}{4 f^2_1 r^4_c}F^2_{tx}}.
\end{eqnarray}

Considering $ \alpha' F_{tx} $ to be small and $ P \gg 1 $ one can expand perturbatively to yield
\begin{eqnarray}
\delta S_{DBI} \approx  \frac{6 \pi^3 T_{D5} \alpha'^2 r_c}{L}\int_0^P d\eta \int d^5x F^2_{tx}+ \cdots ,
\end{eqnarray}
where the metric functions are evaluated taking specific example of $ T_{N_c , P} $ quivers.

In the nonrelativistic limit, one introduces following scaling\footnote{For generic Dp branes the tension is defined as, $ T_{Dp}=\frac{1}{(2 \pi)^p \alpha'^{(p+1)/2}} $. In the TSNC limit, one introduces the nonrelativistic scaling as, $ \alpha' = c^2 \alpha'_{NR} $. This leads to the following relation between the DBI tensions in the relativistic and in the nonrelativistic limit as, $ T_{Dp}=\frac{\tilde{T}_{Dp}}{c^{p+1}} $.}
\begin{eqnarray}
\eta = c^2 \tilde{\eta}~;~\alpha' F_{tx}=c^2 \alpha'_{NR}f_{tx}~;~T_{D5}=\frac{T^{(NR)}_{D5}}{c^6}~;~r_c = c \tilde{r}~;~L= c \ell ,
\end{eqnarray}
which finally yield the nonrelativistic DBI action of the form
\begin{eqnarray}
\delta S^{(NR)}_{DBI}=\frac{6 \pi^3 T^{(NR)}_{D5} \alpha'^2_{NR} \tilde{r}}{c^2 \ell}\int d^5x f^2_{tx}.
\end{eqnarray}

A direct comparison with (\ref{4.7}) reveals that the coupling constant in the nonrelativistic limit of the SCFTs behaves like $ g^2_{QFT} \sim c^2 \rightarrow \infty$. The $1/ c^2 $ dependence is quite intuitive and it stems from the fact that the original location of NS5 branes ($ \eta_i \sim \langle \Phi_i \rangle $) are re-scaled by a factor of $ 1/c^2 $ in the TSNC limit. To summarize, the nonrelativistic limit of $ \mathcal{N}=1 $  quivers emerges as a strongly coupled quantum many body system. 
\subsection{Wilson loops}
\label{sec4.3}
The computation of Wilson loops is carried out starting from the electrostatic description of \cite{Legramandi:2021uds}. The general idea is to consider Wilson loops in the antisymmetric representation of the color gauge group which is realized by considering probe D3 branes extended over $ AdS_2 \subset AdS_6 $ and the two sphere of the internal space.

A direct computation in the relativistic set up yields \cite{Legramandi:2021uds}
\begin{eqnarray}
\label{4.14}
\log \langle \mathcal{W} \rangle =T_{D3}\mathcal{N} \sigma^2 \partial_{\sigma}V,
\end{eqnarray}
where, $ \mathcal{N} = \text{Vol}_{AdS_2}\text{Vol}_{S^2}$.

Let us first estimate the above entity (\ref{4.14}) for $ T_{N_c , P} $ quivers in the large $ P \gg 1 $ limit
\begin{eqnarray}
\sigma^2 \partial_{\sigma}V|_{P \gg 1}=-\frac{\eta N_c P \log 2}{\pi }+\mathcal{O}(1/P).
\end{eqnarray}

Therefore, considering the TSNC limit, one finds
\begin{eqnarray}
\langle \mathcal{W} \rangle|_{c \rightarrow \infty} = e^{-\frac{ \tilde{\eta} \tilde{T}_{D3}\mathcal{N} N_c P \log 2}{\pi c^2}} = 1- \mathcal{O}(c^{-2}),
\end{eqnarray}
which turns out to be a nearly BPS (Wilson) operators whose interpretation in the non-relativistic sector is not clear at the moment.

On a similar note, for $ +_{P, N_c} $ quivers one finds
\begin{eqnarray}
\sigma^2 \partial_{\sigma}V|_{P \gg 1}=- \frac{\eta N_c}{2 \pi}(3+ 2 \log 2)+\mathcal{O}(1/P),
\end{eqnarray}
which in the TSNC limit yields a similar result for Wilson loops
\begin{eqnarray}
\langle \mathcal{W} \rangle|_{c \rightarrow \infty} = 1- \mathcal{O}(c^{-2}).
\end{eqnarray}
\section{Nonrelativistic limits of DGKU solutions}
\label{sec5}
\subsection{Preliminaries}
Let us briefly review the DGKU solution of \cite{DHoker:2016ujz}, which at the first place, was proposed as the holographic set up to describe $ \mathcal{N}=1 $ quivers in 5d. As mentioned previously, the resulting type IIB geometry is a warped product of $ AdS_6$ and $ S^2 $ over a 2d Riemann surface $ \Sigma_{(2)} (\omega , \bar{\omega}) $ parametrised by a pair of complex coordinates $ (\omega , \bar{\omega}) $.

Typically, in their original construction \cite{DHoker:2016ujz}, one introduces a pair of \emph{holomorphic} functions $ \mathcal{A}_{\pm} (\omega)$ whose differentials possess isolated poles on the boundary of $ \Sigma_{(2)} $. They emerge as $ (p ,q) $ five branes with charge density given by the residues at these poles.

Using these holomorphic functions, the type IIB geometry could be expressed as 
\begin{eqnarray}
\label{5.1}
ds^2 = f^2_6 ds^2_{AdS_6}+f^2_2 ds^2_{S^2}+4 \rho^2 d\omega d \bar{\omega},
\end{eqnarray}
where one identifies the warping functions as \cite{Uhlemann:2020bek}
\begin{eqnarray}
f^2_6 = \sqrt{6 \mathcal{G} T}~;~f^2_2 = \frac{1}{9}\sqrt{6 \mathcal{G}}T^{-3/2}~;~\rho^2 = \frac{\kappa^2}{\sqrt{6 \mathcal{G}}}T^{1/2}.
\end{eqnarray}

The metric functions are expressed in terms of local holomorphic functions ($ \mathcal{A}_{\pm} $) \cite{Uhlemann:2020bek}
 \begin{eqnarray}
 \label{5.3}
 \mathcal{G}&=& | \mathcal{A}_{+}|^2 - | \mathcal{A}_{-}|^2 + 2 \text{Re}\mathcal{B}~;~\partial_{\omega}\mathcal{B}= \mathcal{A}_+ \partial_{\omega} \mathcal{A}_- - \mathcal{A}_- \partial_{\omega} \mathcal{A}_+ ,\\
 \kappa^2 & = &- |\partial_{\omega}\mathcal{A}_{+}|^2 +|\partial_{\omega}\mathcal{A}_{-}|^2 ~;~T^2 = 1+ \frac{2 | \partial_{\omega}\mathcal{G}|^2}{3 \kappa^2 \mathcal{G}}.
 \label{5.4}
 \end{eqnarray}
 
 The $ SO(4) $ sector of the dual SCFTs can be realized by wrapping the $ S^3 \subset AdS_6 $. On the other hand, the $ SU(2) $ R symmetry can be realized by wrapping the $ S^2 $ of the internal space. Below, our purpose would be to obtain these metric functions in the nonrelativistic limit and identify the symmetries associated with the dual QFT.
\subsection{Large $ c $ limits of balanced quivers}
Below, we consider some particular examples of $ \mathcal{N}=1 $ quivers with balanced $ SU(N_c) $ color nodes. In these examples, the effective number of flavours is twice the number of colors namely, $ N_f =2N_c $.
\subsubsection{$ T_{N_c} $ quivers}
Let us consider the example of $ T_{N_c} $ quivers. The corresponding holomorphic functions (for the upper half plane) are expressed as \cite{Uhlemann:2020bek}
\begin{eqnarray}
\label{5.7}
\mathcal{A}_{\pm}=\frac{3 N_c}{8 \pi}(\pm \log (\omega -1)+ (\mp 1 - i)\log (\omega + 1)+ i \log (2 \omega)),
\end{eqnarray}
where the poles are clearly located at $ \omega =-1 , 0 ,1  $. Here, $ \omega =1 $ corresponds to $ N_c $ NS5 brane pole. On the other hand, $ \omega =0 $ and $ \omega =-1 $ respectively represent $ N_c $ D5 brane pole and $ N_c $ ($ 1,1 $) five brane pole.

\paragraph{($ p ,q $) five brane web in the nonrelativistic limit.} The poles associated with the differential of the holomorphic function reflect the presence of ($ p ,q $) five branes \cite{Uhlemann:2020bek}. The above notion should get translated in the nonrelativistic limit too. 

In order to define a consistent nonrelativistic limit for holomorphic functions ($ \mathcal{A}_{\pm} $), we rescale the complex coordinate as, $ \omega = c \sqrt{\tilde{\omega}} $ which leads to a holomorphic function in the nonrelativistic limit of the quiver as
\begin{eqnarray}
\mathcal{A}_{\pm}|_{c \rightarrow \infty}=\frac{3 i N_c \log 2}{8 \pi }\mp \frac{\left(\frac{3}{4}\pm \frac{3 i}{8}\right) N_c}{\pi  c \sqrt{\tilde{\omega}} }+\frac{3 i N_c}{16 \pi  c^2 \tilde{\omega} }+ \cdots .
\end{eqnarray}

The differential of the holomorphic function $ \partial_{\omega}\mathcal{A}_{\pm}|_{c \rightarrow \infty} $ exhibits a zero pole at $ \tilde{\omega}\sim 0 $ when expanded upto $ 1/c^2 $ in the large $ c $ limit
\begin{eqnarray}
\partial_{\omega}\mathcal{A}_{\pm}|_{c \rightarrow \infty}=\frac{\left(\pm\frac{3}{4}+\frac{3 i}{8}\right) N_c}{\pi  c^2 \tilde{\omega} }+\mathcal{O}(c^{-3}).
\end{eqnarray}

This is clearly an artefact of the collapsing ($ p,q $) five brane web as a result of the nonrelativistic scaling $ \sim \frac{\omega^2}{c^2} $ with respect to the rescaled coordinate ($ \tilde{\omega} $). For example, the nonrelativistic scaling shifts the previous location ($ \omega =1 $) of NS5 brane pole to $ \tilde{\omega}\sim \frac{1}{c^2}\sim 0 $. The above phenomena precisely mimics the collapsing NS5 brane picture of the electrostatic description using a different ($ \sigma , \eta $) coordinate system (see section \ref{2.2.4}). 

The five brane charges at the pole can be obtained by estimating the residue \cite{Bergman:2018hin}
\begin{eqnarray}
\text{Res}( \partial_{\omega}\mathcal{A}_{\pm})|_{c \rightarrow \infty}=\frac{3\left(\pm 2+i\right) N_c}{8\pi  c^2}=\frac{3(\pm 2\tilde{N}_{NS5}+i \tilde{N}_{D5})}{ 8\pi c^2},
\end{eqnarray}
which shows that the number of NS5 branes is preserved ($\tilde{N}_{NS5} = N_c $) in the nonrelativistic limit. On the other hand, $ \tilde{N}_{D5} = N_c $ counts the net D5 brane charge associated with the zero pole. These numbers precisely indicate towards the conservation of total number of NS5 and D5 branes in the nonrelativistic limit. 

\paragraph{Background geometry.} The background data (in the language of holomorphic functions) can be obtained following the basic definitions those were listed in (\ref{5.3})-(\ref{5.4}).

Using (\ref{5.3}), the equation for the $ \mathcal{B} $ function in the nonrelativistic limit turns out to be
\begin{eqnarray}
\partial_{\omega}\mathcal{B}(\tilde{\omega})|_{c \rightarrow \infty}=-\frac{9 i N_c^2 \log 2}{16 \pi ^2 c^2 \tilde{\omega} }+\mathcal{O}(c^{-3}),
\end{eqnarray}
which upon integration yields
\begin{eqnarray}
\mathcal{B}(\tilde{\omega})|_{c \rightarrow \infty}=\frac{9 i N_c^2 \log 2}{16 \pi ^2 c\sqrt{\tilde{\omega}}}.
\end{eqnarray}

Using these data, it is now straightforward to write down the composite functions (defined at any point $ (\tilde{\omega}) $ of the 2d complex plane) in the nonrelativistic limit as
\begin{eqnarray}
\label{ee5.11}
\mathcal{G}(\tilde{\omega})|_{c \rightarrow \infty}&=&\frac{9 N_c^2 \log 2}{8 \pi ^2 c \left| \tilde{\omega} \right| }\left(\text{Im}\left(\sqrt{\tilde{\omega} }\right)-\left| \tilde{\omega} \right|  \text{Im}\left(\frac{1}{ \sqrt{\tilde{\omega }}} \right)\right)\nonumber\\
&=&\frac{k N^2_c }{c}f_{g}(\tilde{\omega}),
\end{eqnarray}
where $ k $ being some overall constant.

Using (\ref{ee5.11}), the other two functions can be read out as
\begin{eqnarray}
\kappa^2(\tilde{\omega})|_{c \rightarrow \infty} &=& \frac{9 N^2_c f_k(\tilde{\omega})}{16 \pi^2 c^5}~;~f_k(\tilde{\omega})=\text{Im}\left(\frac{ \tilde{\omega}  \left(\left(\tilde{\omega} ^{3/2}\right)^*-\sqrt{\tilde{\omega} } \tilde{\omega} ^*\right)}{ \tilde{\omega}^5}\right)\\
T^2(\tilde{\omega})|_{c \rightarrow \infty}&=&\frac{128 k | \tilde{\omega}|\pi^2 c^2}{27}\frac{f'^2_g}{f_k f_g}.
\label{ee5.13}
\end{eqnarray}

Using the above set of data (\ref{ee5.11})-(\ref{ee5.13}), the metric functions (at any point $ \tilde{\omega} $ of the 2d Riemann surface $ \Sigma_{(2)} $) in the nonrelativistic limit could be formally as
\begin{eqnarray}
\label{ee5.14}
f_6^2 =  g_{6}(\tilde{\omega})~;~f^2_2 =\frac{\sqrt{6}}{9 c^2}g_2 (\tilde{\omega}) ~;~\rho^2 | d \omega |^2 = \frac{g_{\rho} (\tilde{\omega})}{4c^2 |\tilde{\omega}|}| d\tilde{\omega} |^2.
\end{eqnarray}

The details of these functions could be easily read off using the above data (\ref{ee5.11})-(\ref{ee5.13})
\begin{eqnarray}
g_{6}(\tilde{\omega}) \sim \left( \frac{f_g}{f_k}\right)^{1/4} | \tilde{\omega}|^{1/4}\sqrt{f'_g}~;~g_{2}(\tilde{\omega}) \sim \frac{f^{5/4}_g f^{3/4}_k}{f^{'3/2}_{g}|\tilde{\omega}|^{3/4}}~;~g_{\rho} (\tilde{\omega}) \sim  \left( \frac{f_k}{f_g}\right)^{3/4} | \tilde{\omega}|^{1/4}\sqrt{f'_g}.
\end{eqnarray}

Considering $ \Sigma_{(2)} $ to be the upper half plane, it is trivial to see that the pole ($ \tilde{\omega}= 0 $) lies at the origin of the (complex) coordinate system and in particular along  the real axis of the complex plane which is indentified as the boundary of the 2d Riemann surface. 

It is trivial to see that the composite function $ \mathcal{G} $ vanishes identically along the real axis (or the boundary) of the 2d Riemann surface\footnote{On a similar note, one can show that $ \kappa^2|_{\partial \Sigma} =0$ which ensures the boundary conditions of \cite{DHoker:2016ysh} to be valid in the nonrelativistic limit of type IIB solutions.}. In other words, the $ S^2 $ (of the internal manifold) shrinks to zero and as a result the spacetime closes off. This further ensures the geodesic completeness of the TSNC background like in the relativistic scenario \cite{DHoker:2017mds}.

Using (\ref{ee5.14}), the nonrelativistic metric in the large $ c $ limit could be expressed as
\begin{eqnarray}
ds^2 =\frac{1}{c^2}d\tilde{s}^2 = \frac{1}{c^2}(c^2 g_{6}(\tilde{\omega})ds^2_{AdS_6}+\frac{\sqrt{6}}{9}g_2 (\tilde{\omega}) ds^2_{S^2}+\frac{g_{\rho} (\tilde{\omega})}{4 |\tilde{\omega}|}| d\tilde{\omega} |^2).
\end{eqnarray}

Clearly, the longitudinal cloak one forms ($ \tau_{\mu}^A $) are associated with the $ AdS_6 $ factor. For example, the time component of the one form can be expressed as, $ \tau^0_t = \sqrt{g_{6}(\tilde{\omega})} $ and so on. Transverse vielbeins, on the other hand, are associated with $ S^2 $ and the 2d Riemann disc
\begin{eqnarray}
e_{\mu}^a = \frac{6^{1/4}}{3}\sqrt{g_2 (\tilde{\omega}) } ~;~e^{\tilde{a}}_{\tilde{\omega}}=\frac{1}{2}\sqrt{\frac{g_{\rho} (\tilde{\omega})}{|\tilde{\omega}|}}.
\end{eqnarray}
\subsubsection{$ +_{N_c , M} $ quivers}
$ +_{N_c , M} $ quivers are described by the holomorphic functions of the following form \cite{Uhlemann:2020bek}
\begin{eqnarray}
\mathcal{A}_{\pm}=\frac{3}{8 \pi}(i N_c (\log (2 \omega -1)- \log (\omega -1))\pm M (\log (3 \omega -2)-\log \omega)),
\end{eqnarray}
which exhibits $ M $ NS5 brane poles at $ \omega =\lbrace 0 , \frac{2}{3} \rbrace$ and $ N_c $ D5 poles at $ \omega =\lbrace \frac{1}{2} , 1 \rbrace$.

Considering a large $ c $ limit one finds
\begin{eqnarray}
\label{ee5.12}
\mathcal{A}_{\pm}|_{c \rightarrow \infty}= \frac{3 i (N_c \log 2 \mp i M \log 3)}{8 \pi }+\frac{\mp 4 M+3 i N_c}{16 \pi  c \sqrt{\tilde{\omega} }}+\frac{\mp 16 M+27 iN_c}{192 \pi  c^2 \tilde{\omega} }+ \cdots .
\end{eqnarray}

\paragraph{($ p ,q $) five brane web in the nonrelativistic limit.} Clearly, the differential of the holomorphic function (\ref{ee5.12})
\begin{eqnarray}
\partial_{\omega}\mathcal{A}_{\pm}|_{c \rightarrow \infty}=\frac{\pm 4 M-3 i N_c}{16 \pi  c^2 \tilde{\omega }}+ \mathcal{O}(c^{-3})
\end{eqnarray}
exhibits a zero pole in the complex plane which corresponds to a collapsed ($ p ,q $) five brane web like in the previous example. In other words, the nonrelativistic scaling brings all the four ($ p ,q $) five brane poles to the zero of the complex plane. 

The five brane charges could be estimated by computing the residue at the pole \cite{Bergman:2018hin}
\begin{eqnarray}
\text{Res}( \partial_{\omega}  \mathcal{A}_{\pm})|_{c \rightarrow \infty}=\frac{\pm 4 M-3 i N_c}{16 \pi  c^2}=\frac{\pm 4\tilde{N}_{NS5}-3i \tilde{N}_{D5}}{16 \pi c^2}.
\end{eqnarray}

Which clearly indicates that, like in the electrostatic scenario, the number of NS5 branes is preserved $ \tilde{N}_{NS5}= M $. Infact, using the identification, $ M=P $ \cite{Legramandi:2021uds} this is precisely what we identify as the NS5 brane Page charge derived previously using the electrostatic description. On a similar note, $ \tilde{N}_{D5}= N_c $ is the D5 brane charge associated with the pole. 

\paragraph{Background geometry.}  The background data can be obtained pretty much in a way similar to those obtained in the previous example. The $ \mathcal{B} $ function, in the nonrelativistic limit, yields a differential equation of the form
\begin{eqnarray}
\partial_{\omega}\mathcal{B}(\tilde{\omega})|_{c \rightarrow \infty}=-\frac{3 i M N_c \log (432)}{64 \pi ^2 c^2 \tilde{\omega} }+\mathcal{O}(c^{-3}),
\end{eqnarray}
which upon integrating once yields the following solution
\begin{eqnarray}
\mathcal{B}(\tilde{\omega})|_{c \rightarrow \infty}=\frac{3 i M N_c \log (432)}{64 \pi ^2 c \sqrt{\tilde{\omega}} }.
\end{eqnarray}

The rest of the composite functions could be obtained in a straightforward manner
\begin{eqnarray}
\mathcal{G}(\tilde{\omega})|_{c \rightarrow \infty}&=&\frac{3 M N_c \log (432)}{32 \pi ^2 c | \tilde{\omega}|}\left(\text{Im}(\sqrt{\tilde{\omega}}) -|\tilde{\omega}|\text{Im}\left(\frac{1} {\sqrt{\tilde{\omega} }}\right)\right)\nonumber\\ 
&=&\frac{s M N_c }{c}f_{g}(\tilde{\omega}),
\end{eqnarray}
which is of similar structure as that of (\ref{ee5.11}).

The remaining functions can be expressed as
\begin{eqnarray}
\label{E5.25}
\kappa^2 (\tilde{\omega})|_{c \rightarrow \infty} &=& \frac{5 M N_c f_k(\tilde{\omega})}{64 \pi^2 c^5}~;~f_k(\tilde{\omega})=\text{Im}\left(\frac{ (\text{sgn}(\tilde{\omega} )-1) \text{sgn}(\tilde{\omega} )^2}{ \tilde{\omega} ^{5/2}}\right)\\
T^2 (\tilde{\omega})|_{c \rightarrow \infty}&=&\frac{512 s | \tilde{\omega}|\pi^2 c^2}{15}\frac{f'^2_g}{f_k f_g}.
\end{eqnarray}

The rest of the analysis can be carried out precisely in a straightforward manner and the results agree qualitatively to those obtained in the previous section. For example, the cloak one form is given by $ \tau^0_t = \sqrt{g_{6}(\tilde{\omega})} $ with the function $ f_k (\tilde{\omega})$ as given in (\ref{E5.25}).
\subsection{QFT observables}
\paragraph{$(p , q) $ loops.} Let us now calculate the couplings associated with $(p,q)$ strings wrapping the $ AdS_2 \subset AdS_6 $ and thereby taking its nonrelativistic limit. Considering half BPS embeddings those are restricted to the boundary of the disc ($ \partial\Sigma_{(2)} $) one finds \cite{Uhlemann:2020bek}
\begin{eqnarray}
\label{5.10}
S_{(p,q)} = 2 \pi T | (p + i q) (\mathcal{A}_+ - \bar{\mathcal{A}}_-) + (p - i q)(\bar{\mathcal{A}}_+ - \mathcal{A}_-)|.
\end{eqnarray}

In the nonrelativistic limit, one simply makes the replacements $ \mathcal{A}_{\pm} \rightarrow  \mathcal{A}_{\pm} (
\tilde{\omega}_{sol}) $ where $ \tilde{\omega}_{sol} $ is determined considering a nonrelativistic limit of the BPS condition \cite{Uhlemann:2020bek} 
\begin{eqnarray}
\label{5.11}
(p + i q)\partial_{\omega} \mathcal{A}_+|_{c \rightarrow \infty} =(p - i q)\partial_{\omega} \mathcal{A}_-|_{c \rightarrow \infty} .
\end{eqnarray}

Upon solving (\ref{5.11}) for $ T_{N_c} $ quivers, one finds
\begin{eqnarray}
\label{5.12}
\tilde{\omega}_{sol}=\frac{q^2}{c^2 (q-2 p)^2}=\frac{\omega^2_{rel}}{c^2}
\end{eqnarray}
where $ \omega_{rel} $ is the solution obtained in a relativistic five brane web scenario \cite{Uhlemann:2020bek}.

Using (\ref{5.12}), the action (\ref{5.10}) in the nonrelativistic limit turns out to be
\begin{eqnarray}
\label{5.14}
\tilde{S}_{(p,q)}|_{\tilde{\omega}= \tilde{\omega}_{sol}}=-\frac{3N_c\tilde{T}}{2 q c^2}\left |\frac{(q-2 p)^2}{q}+2 q \log 2\right | +\mathcal{O}(c^{-3}).
\end{eqnarray}

Clearly, ($ 1,0 $) strings can be embedded at the (zero) pole of the collapsed ($ p ,q $) five brane web. The expectation value of the corresponding loop operator vanishes
\begin{eqnarray}
\log \langle \mathcal{W}_{(1,0)} (\tilde{z}_F)\rangle |_{c \rightarrow \infty}= - \infty \Rightarrow \langle \mathcal{W}_{(1,0)} (\tilde{z}_F)\rangle |_{c \rightarrow \infty} = 0. 
\end{eqnarray}

On a similar note, ($ 0,1 $) strings are also embedded at the zero pole, $\tilde{\omega}_{sol}\sim  \frac{1}{c^2}\sim 0 $. The expectation value of the associated loop operator is given by
\begin{eqnarray}
\label{5.29}
\log \langle \mathcal{W}_{(0,1)}(\tilde{z}_D) \rangle |_{c \rightarrow \infty} =  -\frac{3 N_c \tilde{T}}{2c^2}(1 + 2 \log 2)\Rightarrow \langle \mathcal{W}_{(0,1)} (\tilde{z}_D)\rangle |_{c \rightarrow \infty}  = 1-\mathcal{O}(c^{-2}).
\end{eqnarray}

Under S-duality the above expectation values must exchange. This is achieved considering the following transformation on the ($ p ,q $) string charges
\begin{eqnarray}
\label{5.30}
\hat{p}= 1 -p ~;~ \hat{q}= 1 - q
\end{eqnarray}
where hat represents string charges in the S-dual picture. 

The corresponding S-dual action is represented as
\begin{eqnarray}
\hat{S}_{(\hat{p}, \hat{q})}= -\frac{3N_c\tilde{T}}{2 (1- \hat{q}) c^2}\left |\frac{(1 + \hat{q}-2 \hat{p})^2}{1 - \hat{q}}+2 (1- \hat{q}) \log 2\right |.
\end{eqnarray}

As for example, the F-strings in the S-dual picture is represented by $ \hat{p}=1 $ and $ \hat{q}=0 $. The expectation value for the corresponding loop operator is $ \langle \hat{\mathcal{W}}_{(1,0)} \rangle   \sim  1$. This is precisely the expectation value associated with D- string operators in (\ref{5.29}). On a similar note, one can show that ($ 1,1 $) strings are mapped into ($ 1 , -1 $) strings under S-duality.

\paragraph{Anti-symmetric Wilson loops and D3 branes.} We revisit the calculation of Wilson loops those were obtained previously using the electrostatic description of $ \mathcal{N}=1 $ quivers.

Wilson loops in the anti-symmetric representation are given by the D3 branes \cite{Uhlemann:2020bek}
\begin{eqnarray}
\log \langle \mathcal{W} (z)\rangle = -\frac{2}{3}T_{D3}\text{Vol}_{AdS_2}\text{Vol}_{S^2}\mathcal{G},
\end{eqnarray}
where $ \mathcal{G} $ is given by (\ref{5.3}).

For $ T_{N_c} $ quivers, it is quite straightforward to evaluate these loops in the nonrelativistic limit. For example, using (\ref{ee5.11}), one finds at any point ($ \tilde{\omega} $) of the upper half plane
\begin{eqnarray}
\log \langle \mathcal{W}(\tilde{z}) \rangle |_{c \rightarrow \infty}=-\frac{2 k N^2_c f_g (\tilde{\omega})}{3 \pi \alpha'^2_{NR} c^5 } \Rightarrow  \langle \mathcal{W}(\tilde{z}) \rangle|_{c \rightarrow \infty} \simeq 1 - \mathcal{O}(c^{-5}),
\end{eqnarray}
which reveals identical results those were obtained previously using electrostatic approach. Clearly, in the strict nonrelativistic limit, the expectation value of these Wilson loops are one and they are mapped into themselves under the S-duality.

On a similar note, one can show that in the nonrelativistic limit of the ($ p , q $) five brane web, the F1 and D1 charges of the D3 brane at any point ($ \tilde{\omega} $) on $ \Sigma_{(2)} $ turn out to be
\begin{eqnarray}
\label{ee5.38}
N_{F1}|_{c\rightarrow \infty}&=& -\frac{N_c}{\pi  c} \frac{  \sin \left(\frac{\arg (\tilde{\omega} )}{2}\right)}{\sqrt[4]{\tilde{\omega} ^2}} +\mathcal{O}(c^{-3}),\\
N_{D1}|_{c \rightarrow \infty}&=& \frac{2N_c}{\pi  c}\frac{ \sin \left(\frac{\arg (\tilde{\omega} )}{2}\right)}{ \sqrt[4]{\tilde{\omega} ^2}}+\mathcal{O}(c^{-3}),
\label{ee5.39}
\end{eqnarray}
where one needs to take into account only the real values on the r.h.s. of (\ref{ee5.38})-(\ref{ee5.39}).

This also scales the field theory direction associated with the dual quiver \cite{Uhlemann:2020bek} as
\begin{eqnarray}
\label{EE5.39}
 \tilde{z}=\frac{N_{D1}}{N_c}\Big |_{c\rightarrow \infty}=\frac{2}{\pi c} \frac{  \sin \left(\frac{\arg (\tilde{\omega} )}{2}\right)}{ \sqrt[4]{\tilde{\omega} ^2}} +\mathcal{O}(c^{-3}),
\end{eqnarray}
where one needs to take into account the real value on the r.h.s. of (\ref{EE5.39}).

Combining the above two expressions (\ref{ee5.38}) and (\ref{ee5.39}), it is quite instructive to note down the following identity 
\begin{eqnarray}
\label{5.41}
(N_{F1} + i N_{D1})|_{c \rightarrow \infty} = \frac{4}{3} F (\tilde{\omega})(\mathcal{A}_+ + \bar{\mathcal{A}}_-)|_{c \rightarrow \infty},
\end{eqnarray}
where $ F(\tilde{\omega})=\frac{\left| \tilde{\omega } \right|  \sin \left(\frac{\arg (\tilde{\omega} )}{2}\right)}{\sqrt[4]{\tilde{\omega}^2} \text{Im}\left(\sqrt{\tilde{\omega } }\right)} $. The above relation (\ref{5.41}) is the nonrelativistic counterpart of the identity derived in the relativistic theory \cite{Uhlemann:2020bek}.
\subsection{Large $ c $ limits of $ Y_{N_c} $ quivers}
\label{sec 5.4}
Nonrelativistic data for unbalanced quivers like $ Y_{N_c} $ and $ X_{N_c} $ can be obtained following a similar analysis. These are the examples where the central node involves a Chern-Simons term and the effective number of flavour ($ N_f $) is less than $ 2N_c $.

$ Y_{N_c} $ solutions are characterised by holomorphic functions of the form \cite{Uhlemann:2020bek}
\begin{eqnarray}
\mathcal{A}_{\pm}=\frac{3 N_c}{8 \pi}[ (\pm 1 +i)\log (\omega - 1)+(\pm 1 - i)\log (\omega + 1)\mp 2 \log (2 \omega),
\end{eqnarray}
whose poles are clearly located at $ \lbrace 0 , \pm 1\rbrace $.

Considering a large $ c $ limit one finds
\begin{eqnarray}
\mathcal{A}_{\pm}|_{c \rightarrow \infty}==\mp \frac{N_c \log 8}{4 \pi }- \frac{3 i N_c}{4 \pi  c\sqrt{\tilde{\omega}}}\mp \frac{3 N_c}{8 c^2 \pi  \tilde{\omega}}+ \cdots ,
\end{eqnarray}
which clearly exhibits a zero pole at $ \tilde{\omega}\sim 0 $ while taking the following differential
\begin{eqnarray}
\partial_{\omega}\mathcal{A}_{\pm}|_{c \rightarrow \infty}= \frac{3 i N_c}{4 \pi  c^2 \tilde{\omega} }+ \mathcal{O}(c^{-3}).
\end{eqnarray}

Like before, the charges associated at the pole of the ($ p ,q $) five brane web could be estimated by knowing the residue \cite{Bergman:2018hin}
\begin{eqnarray}
\text{Res}( \partial_{\omega}\mathcal{A}_{\pm})|_{c \rightarrow \infty}=\frac{3 i N_c}{4 \pi  c^2}=\frac{3 i \tilde{N}}{4 \pi c^2},
\end{eqnarray}
which shows that $ \tilde{N}= N_c $ is the effective number of branes at the zero pole.

\paragraph{Background geometry.} Following our previous discussion, below we estimate the nonrelativistic background data for $ Y_{N_c} $ quivers. Like before, the first step is to obtain the composite function $ \mathcal{B} $ which obeys a differential equation of the form
\begin{eqnarray}
\partial_{\omega}\mathcal{B}(\tilde{\omega})|_{c \rightarrow \infty}=-\frac{3 i N_c^2 \log 8}{8 \pi ^2 c^2 \tilde{\omega} }+\mathcal{O}(c^{-3}),
\end{eqnarray}
which upon integration yields
\begin{eqnarray}
\label{5.47}
\mathcal{B}(\tilde{\omega})|_{c \rightarrow \infty}=\frac{3 i N_c^2 \log 8}{8 \pi ^2 c\sqrt{\tilde{\omega}}}.
\end{eqnarray}

Using (\ref{5.47}), it is straightforward to obtain the remaining functions
\begin{eqnarray}
\mathcal{G}(\tilde{\omega})|_{c \rightarrow \infty}&=&\frac{3  N_c^2 \log 8}{4 \pi ^2 c}f_g(\tilde{\omega})=\frac{\ell N^2_c}{c}f_g(\tilde{\omega}),\\
\kappa^2 (\tilde{\omega})|_{c \rightarrow \infty} &=& \frac{9 N^2_c f_k(\tilde{\omega})}{8 \pi^2 c^5},\\
T^2 (\tilde{\omega})|_{c \rightarrow \infty}&=&\frac{64 \ell | \tilde{\omega}|\pi^2 c^2}{27}\frac{f'^2_g}{f_k f_g}.
\end{eqnarray}

The rest of the discussion follows qualitatively in a way similar to those for the balanced quivers. For example, the composite function ($ \mathcal{G} $) vanishes along the boundary of the upper half plane thereby ensuring the geodesic completeness of the associated TSNC manifold. The TSNC data for $ Y_{N_c} $ quivers turn out to be identical those of the balanced quivers.

\paragraph{$ (p ,q) $ loops.} Let us now explore the nonrelativistic limit of the BPS condition of \cite{Uhlemann:2020bek} which for the present example reveals a solution of the form
\begin{eqnarray}
\tilde{\omega}_{sol}=\frac{p^2}{c^2 q^2}=\frac{\omega^2_{rel}}{c^2},
\end{eqnarray}
which clearly scales the previous solution of \cite{Uhlemann:2020bek} by a factor of $ 1/c^2 $.

The corresponding ($ p , q $) string action can be expressed as
\begin{eqnarray}
\tilde{S}_{(p,q)}|_{\tilde{\omega}= \tilde{\omega}_{sol}}=-\frac{3N_c\tilde{T}}{ c^2}\left |\frac{ \left(q^2-2 p^2 \log 2\right)}{p}\right |+\mathcal{O}(c^{-3}).
\end{eqnarray}

Clearly, the ($ 0,1 $) strings are located at the zero pole while on the other hand, ($ 1,0 $) strings are embedded away from the pole. Clearly, under S-duality, the expectation values of the ($ 0 ,1  $) and ($ 1 , 0 $) loops are exchanged. This is precisely achieved following (\ref{5.30}), which leads to an S-dual action of the following form
\begin{eqnarray}
\hat{S}_{(\hat{p}, \hat{q})}=-\frac{3N_c\tilde{T}}{ c^2}\left |\frac{ \left((1-\hat{q})^2-2 (1-\hat{p})^2 \log 2\right)}{(1- \hat{p})}\right |.
\end{eqnarray}

\paragraph{Anti-symmetric Wilson loops and D3 branes.} Like in the previous examples, the (antisymmetric) Wilson loop in the nonrelativistic limit can be obtained as
\begin{eqnarray}
\log \langle \mathcal{W}(\tilde{z}) \rangle |_{c \rightarrow \infty}=-\frac{2 \ell N^2_c f_g (\tilde{\omega})}{3 \pi \alpha'^2_{NR} c^5 } \Rightarrow  \langle \mathcal{W}(\tilde{z}) \rangle|_{c \rightarrow \infty} \simeq 1 - \mathcal{O}(c^{-5}),
\end{eqnarray}
which corresponds to D3 branes embedded at any point ($ \tilde{\omega} $) of the upper half plane. 

Finally, we note down the associated F1 and D1 charges of the D3 brane 
\begin{eqnarray}
N_{F1}|_{c\rightarrow \infty}&=& -\frac{2N_c}{\pi  c} \frac{ \sin \left(\frac{\arg (\tilde{\omega} )}{2}\right)}{ \sqrt[4]{\tilde{\omega} ^2}}+\mathcal{O}(c^{-3}),\\
N_{D1}|_{c \rightarrow \infty}&=& 0,
\end{eqnarray}
which clearly reveals a difference when compared with the previous examples. 
\subsection{Large $ c $ limits of $ X_{N_c} $ quivers}
\label{sec5.5}
The corresponding supergravity solution is characterised by a pair of holomorphic functions of the following form \cite{Uhlemann:2020bek}
\begin{eqnarray}
\mathcal{A}_{\pm}=\frac{3 N_c}{8 \pi}[ (\pm 1 +i)(\log (3\omega - 2)- \log \omega )+(\pm 1 - i)(\log (\omega - 1)-  \log (2 \omega -1))],
\end{eqnarray}
whose poles are clearly located at $ \lbrace 0 ,  \frac{2}{3}, \frac{1}{2} , 1\rbrace $.

Considering a large $ c $ limit, one finds
\begin{eqnarray}
\mathcal{A}_{\pm}|_{c\rightarrow \infty}=\mathcal{C}_{\pm}\mp \frac{\left(\frac{7}{16}\pm \frac{i}{16}\right) N_c}{\pi  c \sqrt{\tilde{\omega} }}\mp \frac{\left(\frac{43}{192}\mp \frac{11 i}{192}\right) N_c}{\pi  c^2 \tilde{\omega }}+ \cdots,
\end{eqnarray}
where we define the constants as
\begin{eqnarray}
\mathcal{C}_+ = \frac{3N_c}{8\pi }\left(1+i\right)  (\log 3+i \log 2),\\
\mathcal{C}_- = \frac{3N_c}{8\pi }\left(1+i\right)  (i\log 3+\log 2).
\end{eqnarray}

Taking the differential of the holomorphic function, one finds
\begin{eqnarray}
\partial_{\omega}\mathcal{A}_{\pm}|_{c \rightarrow \infty}= \frac{\left(\frac{7}{16}\pm \frac{i}{16}\right) N_c}{\pi  c^2 \tilde{\omega} }+ \mathcal{O}(c^{-3}).
\end{eqnarray}
which clearly reveals a pole associated with the five brane web located at $ \tilde{\omega}=0 $. As before, the charges associated with the nonrelativistic five brane junction can be estimated through residues evaluated at the pole which shows, $ \tilde{N}= N_c $.

\paragraph{Background geometry.} The TSNC data follow trivially like in the previous examples. Below, we calculate the composite functions those are relevant for our subsequent analysis. 

The $ \mathcal{B} $ function can be expressed as
\begin{eqnarray}
\label{e5.63}
\mathcal{B}(\tilde{\omega})|_{c \rightarrow \infty}=\frac{3 N_c^2 \log \left(\frac{81}{8}\right)}{32 \pi ^2 c \sqrt{\tilde{\omega}} }.
\end{eqnarray}

Using (\ref{e5.63}), the $ \mathcal{G} $ function, in its nonrelativistic limit, can be expressed as
\begin{eqnarray}
\label{5.64}
\mathcal{G}(\tilde{\omega})|_{c \rightarrow \infty}&=&\frac{3 N_c^2  }{16 \pi ^2 c  }\left[\frac{ \log (432) \text{Im}\left(\sqrt{\tilde{\omega} }\right)}{ \left|\tilde{ \omega } \right| } +\log \left(\frac{81}{8}\right)\text{Re}\left( \frac{1}{\sqrt{\tilde{\omega}}}\right) \right]\nonumber\\
&=&\frac{3 N_c^2  }{16 \pi ^2 c  }f_{x}(\tilde{\omega}).
\end{eqnarray}

The remaining functions follow trivially as before which we therefore do not list here. Likewise, one can also obtain the corresponding nonrelativistic geometry and the associated background data ($ \tau_{\mu}^A $, $ e_{\mu}^a $) associated with the 10d background.

\paragraph{$ (p ,q) $ loops.} The BPS equation, in its nonrelativistic limit, yields a solution of the form
\begin{eqnarray}
\label{5.63}
\tilde{\omega}_{sol}=\frac{(43 p+11 q)^2}{36 c^2 (q-7 p)^2}.
\end{eqnarray}

Using (\ref{5.63}), the corresponding ($ p , q $) string action can be expressed as
\begin{eqnarray}
\frac{\tilde{S}_{(p,q)}}{\frac{3N_c\tilde{T}}{ c^2}}=\frac{ \left(p^2 \left(172 \coth ^{-1}5-49\right)-2 p q (-7+54 \log 2+32 \log 3)\right)}{2  (43 p+11 q)}\nonumber\\
-\frac{q^2 (1+22 \log 6)}{2  (43 p+11 q)}+\frac{(q-7 p)}{   \sqrt{\frac{(43 p+11 q)^2}{ (q-7 p)^2}}}+\mathcal{O}(c^{-3}).
\end{eqnarray}

Clearly, the F-strings are embedded at $ \tilde{\omega}_{F}= \frac{1849}{1764 c^2}$. On the other hand, D-strings are embedded at $\tilde{\omega}_{D}= \frac{121}{36 c^2} $. Below, we enumerate the expectation values of the corresponding loop operators in the nonrelativistic limit
\begin{eqnarray}
\log \langle \mathcal{W}_{(1,0)} (\tilde{z}_F)\rangle |_{c \rightarrow \infty}&=& e^{\frac{-3.91151N_c \tilde{T}}{c^2}} \Rightarrow \langle \mathcal{W}_{(1,0)} (\tilde{z}_F)\rangle |_{c \rightarrow \infty} = 1- \mathcal{O}(c^{-2}),\\
\log \langle \mathcal{W}_{(0,1)} (\tilde{z}_D)\rangle |_{c \rightarrow \infty}&=& e^{\frac{-5.23891N_c \tilde{T}}{c^2}} \Rightarrow \langle \mathcal{W}_{(0,1)} (\tilde{z}_D)\rangle |_{c \rightarrow \infty} = 1- \mathcal{O}(c^{-2}).
\end{eqnarray}

Clearly, in the strict nonrelativistic ($ c \rightarrow \infty $) limit, the expectation values for both ($ 1,0 $) and ($ 0,1 $) loop operators are trivially one. This further ensures the invariance of the these loop operators under S-duality. In other words, the S-duality exchanges these operators and the entire $ X_{N_c} $ quiver is mapped into itself.

\paragraph{Anti-symmetric Wilson loops and D3 branes.} Wilson loops, in the antisymmetric representation, follow trivially as before. Using (\ref{5.64}), one finds away from the pole
\begin{eqnarray}
\log \langle \mathcal{W}(\tilde{z}) \rangle |_{c \rightarrow \infty}=-\frac{ N^2_c f_x (\tilde{\omega})}{8 \pi^3 \alpha'^2_{NR} c^5 } \Rightarrow  \langle \mathcal{W}(\tilde{z}) \rangle|_{c \rightarrow \infty} \simeq 1 - \mathcal{O}(c^{-5}).
\end{eqnarray}

Finally, we note down the ($ p ,q $) string charges of the D3 brane in the nonrelativistic limit of the five brane web. A straightforward analysis shows
\begin{eqnarray}
(\tilde{N}_{D1}+\tilde{N}_{F1})|_{c \rightarrow \infty}=\frac{N_c}{\pi  c}\frac{  \sin \left(\frac{\arg (\tilde{\omega} )}{2}\right)}{ \sqrt[4]{\tilde{\omega} ^2}}+\mathcal{O}(c^{-3}),\\
(\tilde{N}_{D1}-\tilde{N}_{F1})|_{c \rightarrow \infty}=\frac{4N_c}{3\pi  c} \frac{  \sin \left(\frac{\arg (\tilde{\omega} )}{2}\right)}{ \sqrt[4]{\tilde{\omega} ^2}} +\mathcal{O}(c^{-3}),
\end{eqnarray}
where one has to take into account only real values on the r.h.s. of the above expressions. 

The  corresponding Wilson loop parameter for the D3 brane can expressed as
\begin{eqnarray}
\tilde{z}=\frac{N_{D1}}{N_c}\Big |_{c\rightarrow \infty}=\frac{7}{6\pi c} \frac{  \sin \left(\frac{\arg (\tilde{\omega} )}{2}\right)}{ \sqrt[4]{\tilde{\omega} ^2}} +\mathcal{O}(c^{-3}).
\end{eqnarray}

\section{Concluding remarks and future directions}
\label{sec6}
The present paper is all about learning lessons on the nonrelativistic limits of 5d $ \mathcal{N}=1 $ SCFTs using torsional String Newton Cartan (TSNC) sigma models those are obtained considering a large $ c \rightarrow \infty $ limit of $ AdS_6 \times S^2 \times \Sigma_{(2)}$ geometry in type IIB supergravity. These nonrelativistic backgrounds posses a nontrivial profile for the cloak one form ($ \tau_{\mu}^A $) which results in a non vanishing torsion 2 form $ \tau_{\mu \nu}=2 \partial_{[\mu}\tau_{\nu]}\neq 0 $.

Our analysis reveals a mutual compatibality between the original DGKU formulation \cite{DHoker:2016ujz} of $ \mathcal{N}=1 $ quivers and the recently proposed electrostatic description \cite{Legramandi:2021uds} of 5d SCFTs in their respective nonrelativistic limits. In particular, the zero pole associated with the differential of the holomorphic functions translate into a corresponding picture of collapsing five brane web near the origin ($ \tilde{\eta} \sim 0 $) of the holographic axis in the electrostatic description. 

On top of this, taking a large $ c \rightarrow \infty $ limit of the relativistic expressions, we are able to show that the total number of five branes is preserved in the nonrelativistic description. These numbers (modulo an overall scaling) are shown to be conserved in both descriptions. In the electrostatic approach, these are precisely given by the Page charges. On the other hand, in the DGKU formalism of \cite{DHoker:2016ujz} these numbers are obtained by estimating the residues associated with the differentials of the holomorphic functions.

The present paper offers a tremendous possibility for further investigations on the nonrelativistic limits of electrostatic descriptions and the associated quiver structure for the unbalanced sector \cite{Uhlemann:2020bek} and compare it with the results those were obtained using holomorphic functions. This include examples like $ X_{N_c} $ and $ Y_{N_c} $ quivers. 

The first step towards understanding these limits would be to identify a potential function $ V(\sigma , \eta) $ for unbalanced quivers and thereby rediscovering the physical phenomena of collapse like in the case of the balanced quivers. This will complete the electrostatic picture and will shed light on the associated S-duality properties for the unbalanced sector.

It would be an interesting future project to decode the TSNC data corresponding to the RR sector and thereby constructing a supersymmetric version of the F-string Galilei algebra in the nonrelativistic limits of $ \mathcal{N}=1 $ backgrounds. Another interesting direction would be to explore the nonrelativistic limits of type IIB supergravity equations those were outlined in the introduction and find their compatibility with the beta function calculations.

We leave all these issues for future investigations.\\
{\bf {Acknowledgements :}}
Its a pleasure to thank Carlos Nunez for several useful comments on the draft.
The author is indebted to the authorities of IIT Roorkee for their unconditional support towards researches in basic sciences. The author acknowledges The Royal Society, UK for financial assistance and the Grant (No. SRG/2020/000088) received from The Science and Engineering Research Board (SERB), India. Finally, I wish to thank Dr. Debashree Chowdhury for helping me in drawing those figures in the manuscript.
\appendix
\section{A note on RR fields and matching relations in the $ c \rightarrow \infty $ limit}
\label{appen A}
We consider the nonrelativistic limit of rest of the fields in the supergravity solution those are listed in (\ref{e2.4})-(\ref{e2.11}). These include the RR -sector and the dilaton of the NS sector.

Considering $ T_{N_c ,P} $ quivers, one finds the following expressions near $ \sigma \sim 0 $
\begin{eqnarray}
e^{-2\phi}|_{c \rightarrow \infty}&=&\frac{13 \pi ^2  \eta ^2}{72 P^2}+ \mathcal{O}(N_c c^{-4}),\\
\mathcal{C}_{0}|_{c \rightarrow \infty}&=&-\frac{ \pi ^3 N_c \eta ^2 \sigma}{72  c^2 P^3 \log (2)},\\
\mathcal{C}_{2}|_{c \rightarrow \infty}&=&-\frac{\pi ^2  N_c\eta^3\chi }{P^2 \log (64)}d \chi \wedge d\xi +\mathcal{O}(c^{-4}),
\end{eqnarray}
where we scale, $ P \rightarrow c^2 P $ and remove tildes on the r.h.s. of the above set of relations.

As we mentioned previously, a complete understanding of these nonrelativistic limits in terms of the TSNC data is not known yet. In other words, an analogous relation to that of (\ref{e2.26}) is yet to be settled down.

Under S-duality, these backgrounds transform into a different $ AdS_6 $ vacuum in its nonrelativistic limit. Our purpose would be to identify this new type IIB vacuum obtained via S-duality. To this end, we choose to work with the type IIB background of \cite{Apruzzi:2018cvq} as presented by authors in \cite{Legramandi:2021uds}.

Taking a nonrelativistic limit, one finds (modulo an overall scaling that may be absorbed into the definition of the background fluxes)
\begin{eqnarray}
\hat{B}_2|_{c \rightarrow \infty} =-\mathcal{C}_{2}|_{c \rightarrow \infty}~;~\hat{\mathcal{C}}_2|_{c \rightarrow \infty} = B_2|_{c \rightarrow \infty},
\end{eqnarray}
where hatted fields correspond to a nonrelativistic limit of type IIB solutions of \cite{Apruzzi:2018cvq}.

\paragraph{Matching relations in the nonrelativistic limit.} We now discuss the nonrelativistic limits of the matching relations those were presented in \cite{Legramandi:2021uds}. These matching relations precisely serve as the dictionary between the two parallel descriptions of $ \mathcal{N}=1 $ quivers. 

For $ T_{N_c , P} $ quivers, the mapping is given by the following relation 
\begin{eqnarray}
\omega = \coth \left( \frac{2 \pi z}{9 N_c}\right),
\end{eqnarray}
where $ z = \sigma - i \eta $ \cite{Legramandi:2021uds}.

In the nonrelativistic limit, we introduce the scaling relation of the following form
\begin{eqnarray}
z = \frac{1}{ c \tilde{z}},
\end{eqnarray}
which in the large $ c $ limit yields a map of the following form
\begin{eqnarray}
\tilde{\omega}=\frac{81 N^2_c \tilde{z}^2}{4 \pi^2 }+\mathcal{O}(c^{-2}).
\end{eqnarray}

Clearly, the zero pole of the holomorphic function corresponds to setting $ \tilde{z}=0 $. Considering an expansion near $ \tilde{\sigma }=0 $, corresponds to setting $ \tilde{\eta}=0 $. This is precisely the origin of the holographic axis in the nonrelativistic limit of the electrostatic description.

On a similar note, for $ +_{M , N_c} $ quivers one finds \cite{Legramandi:2021uds}
\begin{eqnarray}
\omega = \frac{2}{3}\left(1+\frac{1}{3 e^{4 \pi z/9M}-1} \right).
\end{eqnarray}

Considering a nonrelativistic expansion of the form
\begin{eqnarray}
z=(\tilde{z} - z_0)/c ~;~z_0 = -\frac{9 M c}{2 \pi}
\end{eqnarray}
one finds the following mapping between the variables in two descriptions
\begin{eqnarray}
\tilde{\omega}\approx \frac{4 \pi^2 \tilde{z}^2}{81 M^2 c^4}.
\end{eqnarray}
\section{Detailed expressions for the metric functions in Region III}
\label{appendixA}
The individual functions $ a_i(\eta) $ and $ b_i(\eta) $ read as
\begin{eqnarray}
a_1 (\eta)&=&-\left(\eta ^2+1\right) \log \left(\frac{\eta +1}{\eta -1}\right)+\eta  \left(-2 \log \left(\eta ^2-1\right)+6+\log 16\right)-4 \eta  \log \left(\frac{\pi }{P}\right),\\
b_1(\eta)&=&\log \left(\frac{\eta +1}{\eta -1}\right),
\end{eqnarray}
\begin{eqnarray}
\frac{a_2 (\eta)}{\sqrt{3} \pi}&=&  \left(\left(\eta ^2+1\right) \log \left(\frac{\eta +1}{\eta -1}\right)-2 \eta  \left(-\log \left(\eta ^2-1\right)+3+\log 4\right)+4 \eta  \log \left(\frac{\pi }{P}\right)\right)^2,\\
b_2(\eta)&=&c_2(\eta)\sqrt{\frac{-2 \left(\eta ^2+1\right) \log \left(\frac{\eta +1}{\eta -1}\right)+4 \eta  \left(-\log \left(\eta ^2-1\right)+3+\log 4\right)-8 \eta  \log \left(\frac{\pi }{P}\right)}{\log \left(\frac{\eta +1}{\eta -1}\right)}},\\
c_2(\eta)&=&-2 \left(\eta ^2+3\right) \log ^2\left(\frac{\eta +1}{\eta -1}\right)+\left(-2 \log \left(\eta ^2-1\right)+4+\log 16\right)^2+16 \log ^2\left(\frac{\pi }{P}\right),
\end{eqnarray}
\begin{eqnarray}
a_3(\eta)=\sqrt{\frac{3}{2}} \pi  \log \left(\frac{\eta +1}{\eta -1}\right)~~~~~~~~~~~~~~~~~~~~~~~~~~~~~~~~~~~~~~~~~~~~~~~~~~~~~~~~~~~~~~~~~~~~\nonumber\\
\times \sqrt{\frac{-\left(\eta ^2+1\right) \log \left(\frac{\eta +1}{\eta -1}\right)+\eta  \left(-2 \log \left(\eta ^2-1\right)+6+\log 16\right)-4 \eta  \log \left(\frac{\pi }{P}\right)}{\log \left(\frac{\eta +1}{\eta -1}\right)}},
\end{eqnarray}
\begin{eqnarray}
b_3(\eta)=\left(\eta ^2+1\right) \log \left(\frac{\eta +1}{\eta -1}\right)-2 \eta  \left(-\log \left(\eta ^2-1\right)+3+\log 4\right)+4 \eta  \log \left(\frac{\pi }{P}\right).
\end{eqnarray}


\begin{thebibliography}{99}
\bibitem{Gomis:2000bd}
J.~Gomis and H.~Ooguri,
``Nonrelativistic closed string theory,''
J. Math. Phys. \textbf{42}, 3127-3151 (2001)
doi:10.1063/1.1372697
[arXiv:hep-th/0009181 [hep-th]].

\bibitem{Gomis:2005pg}
J.~Gomis, J.~Gomis and K.~Kamimura,
``Non-relativistic superstrings: A New soluble sector of AdS(5) x S**5,''
JHEP \textbf{12}, 024 (2005)
doi:10.1088/1126-6708/2005/12/024
[arXiv:hep-th/0507036 [hep-th]].

\bibitem{Bergshoeff:2015uaa}
E.~Bergshoeff, J.~Rosseel and T.~Zojer,
``Newton\textendash{}Cartan (super)gravity as a non-relativistic limit,''
Class. Quant. Grav. \textbf{32}, no.20, 205003 (2015)
doi:10.1088/0264-9381/32/20/205003
[arXiv:1505.02095 [hep-th]].

\bibitem{Bergshoeff:2021bmc}
E.~A.~Bergshoeff, J.~Lahnsteiner, L.~Romano, J.~Rosseel and C.~Simsek,
``A non-relativistic limit of NS-NS gravity,''
JHEP \textbf{06}, 021 (2021)
doi:10.1007/JHEP06(2021)021
[arXiv:2102.06974 [hep-th]].

\bibitem{Bergshoeff:2019pij}
E.~A.~Bergshoeff, J.~Gomis, J.~Rosseel, C.~Simsek and Z.~Yan,
``String Theory and String Newton-Cartan Geometry,''
J. Phys. A \textbf{53}, no.1, 014001 (2020)
doi:10.1088/1751-8121/ab56e9
[arXiv:1907.10668 [hep-th]].

\bibitem{Bergshoeff:2018yvt}
E.~Bergshoeff, J.~Gomis and Z.~Yan,
``Nonrelativistic String Theory and T-Duality,''
JHEP \textbf{11}, 133 (2018)
doi:10.1007/JHEP11(2018)133
[arXiv:1806.06071 [hep-th]].

\bibitem{Harmark:2017rpg}
T.~Harmark, J.~Hartong and N.~A.~Obers,
``Nonrelativistic strings and limits of the AdS/CFT correspondence,''
Phys. Rev. D \textbf{96}, no.8, 086019 (2017)
doi:10.1103/PhysRevD.96.086019
[arXiv:1705.03535 [hep-th]].

\bibitem{Harmark:2018cdl}
T.~Harmark, J.~Hartong, L.~Menculini, N.~A.~Obers and Z.~Yan,
``Strings with Non-Relativistic Conformal Symmetry and Limits of the AdS/CFT Correspondence,''
JHEP \textbf{11}, 190 (2018)
doi:10.1007/JHEP11(2018)190
[arXiv:1810.05560 [hep-th]].

\bibitem{Gomis:2020izd}
J.~Gomis, Z.~Yan and M.~Yu,
``T-Duality in Nonrelativistic Open String Theory,''
JHEP \textbf{02}, 087 (2021)
doi:10.1007/JHEP02(2021)087
[arXiv:2008.05493 [hep-th]].

\bibitem{Hartong:2021ekg}
J.~Hartong and E.~Have,
``Nonrelativistic Expansion of Closed Bosonic Strings,''
Phys. Rev. Lett. \textbf{128}, no.2, 021602 (2022)
doi:10.1103/PhysRevLett.128.021602
[arXiv:2107.00023 [hep-th]].

\bibitem{Harmark:2019upf}
T.~Harmark, J.~Hartong, L.~Menculini, N.~A.~Obers and G.~Oling,
``Relating non-relativistic string theories,''
JHEP \textbf{11}, 071 (2019)
doi:10.1007/JHEP11(2019)071
[arXiv:1907.01663 [hep-th]].

\bibitem{Gallegos:2019icg}
A.~D.~Gallegos, U.~G\"ursoy and N.~Zinnato,
``Torsional Newton Cartan gravity from non-relativistic strings,''
JHEP \textbf{09}, 172 (2020)
doi:10.1007/JHEP09(2020)172
[arXiv:1906.01607 [hep-th]].

\bibitem{Gomis:2019zyu}
J.~Gomis, J.~Oh and Z.~Yan,
``Nonrelativistic String Theory in Background Fields,''
JHEP \textbf{10}, 101 (2019)
doi:10.1007/JHEP10(2019)101
[arXiv:1905.07315 [hep-th]].

\bibitem{Bidussi:2021ujm}
L.~Bidussi, T.~Harmark, J.~Hartong, N.~A.~Obers and G.~Oling,
``Torsional string Newton-Cartan geometry for non-relativistic strings,''
JHEP \textbf{02}, 116 (2022)
doi:10.1007/JHEP02(2022)116
[arXiv:2107.00642 [hep-th]].

\bibitem{Yan:2021lbe}
Z.~Yan,
``Torsional deformation of nonrelativistic string theory,''
JHEP \textbf{09}, 035 (2021)
doi:10.1007/JHEP09(2021)035
[arXiv:2106.10021 [hep-th]].

\bibitem{Maldacena:1997re}
J.~M.~Maldacena,
``The Large N limit of superconformal field theories and supergravity,''
Adv. Theor. Math. Phys. \textbf{2}, 231-252 (1998)
doi:10.1023/A:1026654312961
[arXiv:hep-th/9711200 [hep-th]].

\bibitem{Seiberg:1996bd}
N.~Seiberg,
``Five-dimensional SUSY field theories, nontrivial fixed points and string dynamics,''
Phys. Lett. B \textbf{388}, 753-760 (1996)
doi:10.1016/S0370-2693(96)01215-4
[arXiv:hep-th/9608111 [hep-th]].

\bibitem{DHoker:2016ujz}
E.~D'Hoker, M.~Gutperle, A.~Karch and C.~F.~Uhlemann,
``Warped $AdS_6\times S^2$ in Type IIB supergravity I: Local solutions,''
JHEP \textbf{08}, 046 (2016)
doi:10.1007/JHEP08(2016)046
[arXiv:1606.01254 [hep-th]].

\bibitem{DHoker:2016ysh}
E.~D'Hoker, M.~Gutperle and C.~F.~Uhlemann,
``Holographic duals for five-dimensional superconformal quantum field theories,''
Phys. Rev. Lett. \textbf{118}, no.10, 101601 (2017)
doi:10.1103/PhysRevLett.118.101601
[arXiv:1611.09411 [hep-th]].

\bibitem{DHoker:2017mds}
E.~D'Hoker, M.~Gutperle and C.~F.~Uhlemann,
``Warped $AdS_6\times S^2$ in Type IIB supergravity II: Global solutions and five-brane webs,''
JHEP \textbf{05}, 131 (2017)
doi:10.1007/JHEP05(2017)131
[arXiv:1703.08186 [hep-th]].

\bibitem{DHoker:2017zwj}
E.~D'Hoker, M.~Gutperle and C.~F.~Uhlemann,
``Warped $AdS_6\times S^2$ in Type IIB supergravity III: Global solutions with seven-branes,''
JHEP \textbf{11}, 200 (2017)
doi:10.1007/JHEP11(2017)200
[arXiv:1706.00433 [hep-th]].

\bibitem{Fluder:2020pym}
M.~Fluder and C.~F.~Uhlemann,
``Evidence for a 5d F-theorem,''
JHEP \textbf{02}, 192 (2021)
doi:10.1007/JHEP02(2021)192
[arXiv:2011.00006 [hep-th]].

\bibitem{Gutperle:2020rty}
M.~Gutperle and C.~F.~Uhlemann,
``Surface defects in holographic 5d SCFTs,''
JHEP \textbf{04}, 134 (2021)
doi:10.1007/JHEP04(2021)134
[arXiv:2012.14547 [hep-th]].

\bibitem{Uhlemann:2020bek}
C.~F.~Uhlemann,
``Wilson loops in 5d long quiver gauge theories,''
JHEP \textbf{09}, 145 (2020)
doi:10.1007/JHEP09(2020)145
[arXiv:2006.01142 [hep-th]].

\bibitem{Uhlemann:2019ypp}
C.~F.~Uhlemann,
``Exact results for 5d SCFTs of long quiver type,''
JHEP \textbf{11}, 072 (2019)
doi:10.1007/JHEP11(2019)072
[arXiv:1909.01369 [hep-th]].

\bibitem{Uhlemann:2019lge}
C.~F.~Uhlemann,
``AdS$_6$/CFT$_5$ with O7-planes,''
JHEP \textbf{04}, 113 (2020)
doi:10.1007/JHEP04(2020)113
[arXiv:1912.09716 [hep-th]].

\bibitem{Bergman:2018hin}
O.~Bergman, D.~Rodr\'\i{}guez-G\'omez and C.~F.~Uhlemann,
``Testing AdS$_{6}$/CFT$_{5}$ in Type IIB with stringy operators,''
JHEP \textbf{08}, 127 (2018)
doi:10.1007/JHEP08(2018)127
[arXiv:1806.07898 [hep-th]].

\bibitem{Legramandi:2021uds}
A.~Legramandi and C.~Nunez,
``Electrostatic description of five-dimensional SCFTs,''
Nucl. Phys. B \textbf{974}, 115630 (2022)
doi:10.1016/j.nuclphysb.2021.115630
[arXiv:2104.11240 [hep-th]].

\bibitem{Roychowdhury:2021jqt}
D.~Roychowdhury,
``Non-integrability for $ \mathcal{N}=1 $ SCFTs in $ 5d $,''
Phys. Rev. D \textbf{104}, no.8, 086010 (2021)
doi:10.1103/PhysRevD.104.086010
[arXiv:2106.10646 [hep-th]].

\bibitem{Apruzzi:2018cvq}
F.~Apruzzi, J.~C.~Geipel, A.~Legramandi, N.~T.~Macpherson and M.~Zagermann,
``Minkowski$_4$ $\times$ $S^2$ solutions of IIB supergravity,''
Fortsch. Phys. \textbf{66}, no.3, 1800006 (2018)
doi:10.1002/prop.201800006
[arXiv:1801.00800 [hep-th]].

\bibitem{Alvarez:1994wj}
E.~Alvarez, L.~Alvarez-Gaume and Y.~Lozano,
``A Canonical approach to duality transformations,''
Phys. Lett. B \textbf{336}, 183-189 (1994)
doi:10.1016/0370-2693(94)00982-1
[arXiv:hep-th/9406206 [hep-th]].

\end{thebibliography}
\end{document}